\documentclass[twocolumn]{aastex631}  
\usepackage{amsmath}
\usepackage{graphicx}
\usepackage{hyperref}
\usepackage{natbib}

\graphicspath{{figures/}}

\newcommand\HST{\textit{HST}}
\newcommand\JWST{\textit{JWST}}

\newcommand{\U}[2]{$\mathcal{U}(#1, #2)$}
\newcommand{\D}[2]{$\mathcal{D}(#1, #2)$}
\newcommand\R[1]{#1}
\defcitealias{Niraula2022}{N22}

\shorttitle{Towards Robust Stellar Contamination Corrections in JWST Spectra}
\shortauthors{Rackham \& de Wit}
\graphicspath{{./}{figures/}}

\begin{document}



\title{Toward robust corrections for stellar contamination in JWST exoplanet transmission spectra}

\author[0000-0002-3627-1676]{Benjamin V.\ Rackham}
\altaffiliation{51 Pegasi b Fellow}
\affiliation{Department of Earth, Atmospheric and Planetary Sciences, Massachusetts Institute of Technology, 77 Massachusetts Avenue, Cambridge, MA 02139, USA}
\affiliation{Kavli Institute for Astrophysics and Space Research, Massachusetts Institute of Technology, Cambridge, MA 02139, USA}

\author[0000-0003-2415-2191]{Julien de Wit}
\affiliation{Department of Earth, Atmospheric and Planetary Sciences, Massachusetts Institute of Technology, 77 Massachusetts Avenue, Cambridge, MA 02139, USA}
\affiliation{Kavli Institute for Astrophysics and Space Research, Massachusetts Institute of Technology, Cambridge, MA 02139, USA}




\begin{abstract}
Transmission spectroscopy is still the preferred characterization technique for exoplanet atmospheres,  although it presents unique challenges that translate into characterization bottlenecks when robust mitigation strategies are missing. 
Stellar contamination is one such challenge that can overpower the planetary signal by up to an order of magnitude, and thus not accounting for \R{it} can lead to significant biases in the derived atmospheric properties. 
Yet \R{this} accounting may not be straightforward, as important discrepancies exist between state-of-the-art stellar models and measured spectra and between models themselves. 
Here we explore the extent to which stellar models can be used to reliably correct for stellar contamination and yield a planet's uncontaminated transmission spectrum. 
We find that 
discrepancies between stellar models can \R{significantly contribute to} the noise budget of \JWST{} transmission spectra of planets around stars with heterogeneous photospheres,
the true number of unique photospheric spectral components and their properties can only be accurately retrieved when the stellar models have sufficient fidelity, and 
under such optimistic circumstances the contribution of stellar contamination to the noise budget of a transmission spectrum is considerably below that of the photon noise for the standard transit observation setup. 
Therefore, we \R{advocate for further development of}
model spectra of stars and their active regions in a data-driven manner, 
empirical approaches for deriving spectra of photospheric components using the observatories with which the atmospheric explorations are carried out\R{, and}
\R{analysis techniques accounting for multimodal posterior distributions for photospheric parameters of interest, which will be increasingly revealed by precise \JWST{} measurements.}
\end{abstract}



\section{Introduction} \label{sec:intro}

Transmission spectroscopy, the multiwavelength study of the shadows cast by transiting exoplanets \citep[e.g.,][]{Seager2000, Brown2001}, provides a powerful tool for constraining the physical structure and chemical composition of exoplanet atmospheres, as recently demonstrated by the \JWST{} Early Release Science observations of WASP-39b \citep{JTECERST2023, Ahrer2023, Alderson2023, Feinstein2023, Rustamkulov2023_ERS}.
However, the transmission spectrum only contains information related to the wavelength-dependent opacity of a planet’s atmosphere alone when the stellar disk is limb darkened but otherwise featureless.
For stars with notable coverage of photospheric heterogeneities like spots and faculae, the difference in the hemisphere-averaged emission spectrum of the star and the transit-chord-averaged one imprints features in the transmission spectrum \citep[e.g.,][]{Sing2011, McCullough2014, Rackham2017}, a phenomenon dubbed the transit light source (TLS) effect \citep{Rackham2018, Rackham2019}.
Active FGK stars and nearly all M dwarfs are expected to produce detectable TLS, or ``stellar contamination,'' signals in precise transmission spectra.
\R{The imprints of the TLS effect are most evident at shorter (visible and ultraviolet) wavelengths, where the contrast between the quiescent photosphere, spots, and faculae is greatest \citep{Rackham2018, Rackham2019}, though impacts on near-infrared observations are possible for active stars as well \citep{ZhangZhanbo2018, Lim2023}.}

\R{
In this context, recent work has explored the potential to identify stellar contamination signals in transmission spectra and disentangle them from signals arising from the planetary atmosphere.
\citet{Pinhas2018} presented a framework for jointly constraining stellar and planetary spectral signals in transmission spectra using nested sampling to assess the evidence for models that include stellar contamination.
Applying this framework to a sample of hot giant exoplanets observed by the \textit{Hubble Space Telescope} (\HST{}), they find evidence for stellar contamination in two of nine spectra.
Looking forward to \JWST{}, \citet{Iyer2020} explored the potential impact of stellar contamination on \JWST{} transmission spectra of sub-Neptunes transiting M dwarfs, finding that biased inferences of planetary properties are possible when ignoring stellar heterogeneity corrections or applying stellar models that have limited fidelity with respect to the true stellar spectrum, i.e., models that do not accurately match the data.
\citet{Iyer2020} also tested the utility of using the out-of-transit stellar spectrum to constrain stellar photospheric heterogeneity and find that this approach offers no improvement over an analysis of the in-transit data in the case where the models have high fidelity with respect to the true stellar spectrum and the number of distinct spectral components (e.g., quiescent photosphere, spots, faculae) is not in question.
}

\R{At the same time, other recent work has focused on constraining and mitigating} for the heterogeneity of the stellar disk at the time of transit by leveraging the stellar spectrum collected during the out-of-transit baseline.
Specifically, \R{in this approach} the temperatures and filling factors of the different photospheric components are constrained to \R{enable corrections} for their contributions to the joint in-transit spectrum \citep[e.g.,][]{ZhangZhanbo2018,Wakeford2019,Garcia2022}.
Early on, these mitigation studies revealed two bottlenecks.
First, fitting host star spectra with the precisions afforded by space-based platforms is a challenge for current models, especially for late M dwarfs such as TRAPPIST-1 \citep{Gillon2016,Gillon2017}.
\citet{ZhangZhanbo2018} showed that the uncertainties on the HST/WFC3/G141 spectra of TRAPPIST-1 need to be inflated by factors of $\sim$23 to produce adequate fits with respect to stellar models.
The subsequent studies of \citet{Wakeford2019} and \citet{Garcia2022} yielded consistent challenges, which are expected to worsen in the \JWST{} era following a substantial increase in precision \citep[see recent review from][]{Rackham2023}.
Second, with the current data quality and model fidelity, an ensemble of models may fit an out-of-transit spectrum equally well, leading to a range of corrected atmospheric spectra with a scatter many times larger than the photon noise (see, e.g., Fig.~6 from \citealt{Wakeford2019} and Fig.~7 from \citealt{Garcia2022}).

As stellar contamination can overpower the planetary signal by up to an order of magnitude \citep{Rackham2018}, not accounting for it when performing atmospheric retrieval can lead to important biases in inferred planetary atmospheric parameters \citep{Iyer2020}.
A zeroth-order mitigation strategy to account for the imperfections of stellar models and avoid biases in the corrected planetary spectra is thus to inflate the uncertainties of the stellar spectra \citep[e.g.,][]{ZhangZhanbo2018}, thereby decreasing the precision of planetary inferences.
However, the optimal study of exoplanet atmospheres with current facilities demands refined mitigation approaches that can harness the precision of these observations to reduce biases and uncertainties as much as possible.

Here we explore the limits of using baseline out-of-transit observations to infer the photospheric properties of cool stars and mitigate for stellar contamination in transmission spectra, with a particular consideration for the fidelity of current stellar models.
Our analysis complements that recently conducted for opacity models by \citet[][hereafter N22]{Niraula2022}.
\R{Our work builds on previous studies of heterogeneity constraints enabled by out-of-transit stellar spectra by considering K and M spectral types with a range of activity levels and, presenting an analysis framework that assesses the suitability of models with one to four spectral components, uses Bayesian model averaging to derive the final spectral model and relevant corrections to transmission spectra, and propagates the additional uncertainty due to limited model fidelity onto the stellar contamination correction.}
We first investigate the utility of out-of-transit \JWST{} spectra for identifying a complex photosphere with multiple spectral components and whether inferences are limited by the data quality at hand or the fidelity of stellar spectral models.
If the latter, this means that new stellar models (theoretical or empirical) may help us move toward a photon-noise-dominated regime.
Then, setting aside model fidelity, we assess whether this approach permits inferences of photospheric parameters that are accurate and precise enough to reduce biases in transmission spectra.
Finally, we evaluate the contribution of stellar contamination to the total noise and how this scales with the ratio of the out-of-transit to in-transit observations.

Note that we focus in this paper on configurations in which heterogeneities are present but not occulted by the transiting exoplanet.
Mitigation strategies for occulted active regions can be found in other studies \citep[e.g.,][]{Fu2022}.
\R{
Relevant to our analysis here, we note that when active regions are occulted by the transiting planet, one can constrain the temperature of the occulted heterogeneity.
This constraint can then be used as a prior on the out-of-transit spectral analysis to infer the total covering fraction of the heterogeneity and, in turn, the impact of any unocculted features.
}

This paper is organized as follows.
\autoref{sec:data} presents our approach for generating the synthetic datasets for analysis.
\autoref{sec:retrievals} details our retrieval approach for inferring constraints from simulated out-of-transit stellar spectra.
\autoref{sec:results} shares our results, and \autoref{sec:discussion} summarizes our findings while placing them in the larger context of \JWST{} observations.

\section{Data synthesis} \label{sec:data}

In order to explore the ability of current stellar models to support the reliable correction of stellar contamination in \JWST{} exoplanet transmission spectra, we follow a sensitivity analysis similar to that introduced in \citetalias{Niraula2022} for opacity models.
We explore two systems and five levels of heterogeneities in our sensitivity analysis, which we describe in the following.

\subsection{Properties of the Synthetic Systems}
\label{sec:system_parameters}

We adopt synthetic systems similar to those introduced in \citetalias{Niraula2022} as representative examples of planets that would be high-priority targets for \JWST{}.
These correspond to an Earth-sized planet around an M dwarf star and a Jupiter-sized planet around a K dwarf star. 
\R{We note that we adopt these planetary parameters and transmission spectra solely in order to simulate precisions for nominal cases.
The specifics of these systems do not impact our analysis because we are strictly interested in inferences of stellar parameters and associated contamination signals that derive from an analysis of the out-of-transit stellar spectra.}

The warm Jupiter has a mass of $1\,M_\mathrm{jup}$, radius of $1\,R_\mathrm{jup}$, a reference temperature of 500\,K, and a transit duration of 5.80\,hr.
The super-Earth has a mass of $1\,M_\mathrm{\earth}$, radius of $1\,R_\mathrm{\earth}$, a reference temperature of 300\,K, and a transit duration of 1.00\,hr.
The details of the atmospheric model of each planet, given in Table\,2 of \citetalias{Niraula2022}, are not important for this analysis, as we are interested instead in the impact of the host stars.

The K dwarf has an effective temperature of $T_\mathrm{eff} = 5270$\,K, a stellar mass of $M_s = 0.88\,M_\sun$, and a stellar radius of $R_s = 0.813\,R_\sun$, parameters that correspond to a K0 dwarf \citep{Pecaut2013}.
For the M dwarf, $T_\mathrm{eff} = 2810\,K$, $M_s = 0.102\,M_\sun$, and $R_s = 0.137\,R_\sun$, corresponding to an M6 dwarf \citep{Pecaut2013}.
In both cases, we consider solar-metallicity stars ($\mathrm{[Fe/H]} = 0.0$).
We also adopt a brightness of $J=11$ for both host stars, \R{corresponding to} distances of 191 and 20.5\,pc for the K0 and M6 stars, respectively.

\subsection{Synthetic Cases of Photospheric Heterogeneity} \label{sec:stellar_heterogeneity}

For each host star, we consider five photospheric heterogeneity scenarios, detailed in \autoref{tab:cases}.
The first case, \texttt{case 1}, is a quiescent star, for which the quiescent photosphere spectrum is the only spectral component present on the stellar disk.
The next two cases, \texttt{cases 2l} and \texttt{2h}, are for a star with two spectral components, those of the quiescent photosphere and a spot. 
The spot coverage is 1\% in the low-activity case (\texttt{case 2l}) and 5\% in the high-activity case (\texttt{case 2h}).
The last two cases, \texttt{cases 3l} and \texttt{3h}, are for a star with three spectral components, those of the quiescent photosphere, spots, and faculae. 
The coverages of the spots are the same as the previous low-activity and high-activity cases, and the coverages of the faculae are 10\% and 30\% for \texttt{case 3l} and \texttt{case 3h}, respectively.

\begin{deluxetable}{llrr}[t]
\tablecaption{Parameters of the Five Heterogeneity Cases. \label{tab:cases}}
\tablehead{Case & Description & $f_\mathrm{spot}\,(\%)$ & $f_\mathrm{fac}\,(\%)$}
\startdata
\texttt{1} & No activity & 0 & 0 \\
\texttt{2l} & Spots, low activity & 1 & 0 \\
\texttt{2h} & Spots, high activity & 5 & 0 \\
\texttt{3l} & Spots and faculae, low activity & 1 & 10 \\
\texttt{3h} & Spots and faculae, high activity & 5 & 30 \\
\enddata
\end{deluxetable}

For each star, we adopt the effective temperature as the temperature of the quiescent photosphere. 
For the K dwarf, we set the spot and facula temperatures to 3830 and 5380\,K, respectively, following \citet{Rackham2019}. 
For the M dwarf, we set the spot temperature to 86\% of the photospheric temperature (2420\,K) and the facula temperature to 2910\,K, following \citet{Afram2015} and \citet{Rackham2018}, respectively.

We generate the model truth of the (out-of-transit) stellar spectrum as the linear combination of the constituent spectra weighted by their filling factors.
We do not assume any specific position on the stellar disk for the spots and faculae besides that they are present outside of the transit chord and thus undetectable via crossing events \citep[e.g.,][]{Fu2022}.
As a result, we take the component spectra to be representative of spots at all positions and neglect the impact of limb darkening.

\subsection{Stellar Spectral Models} 

\begin{figure*}[htbp!]
    \centering
    \includegraphics[width=\textwidth]{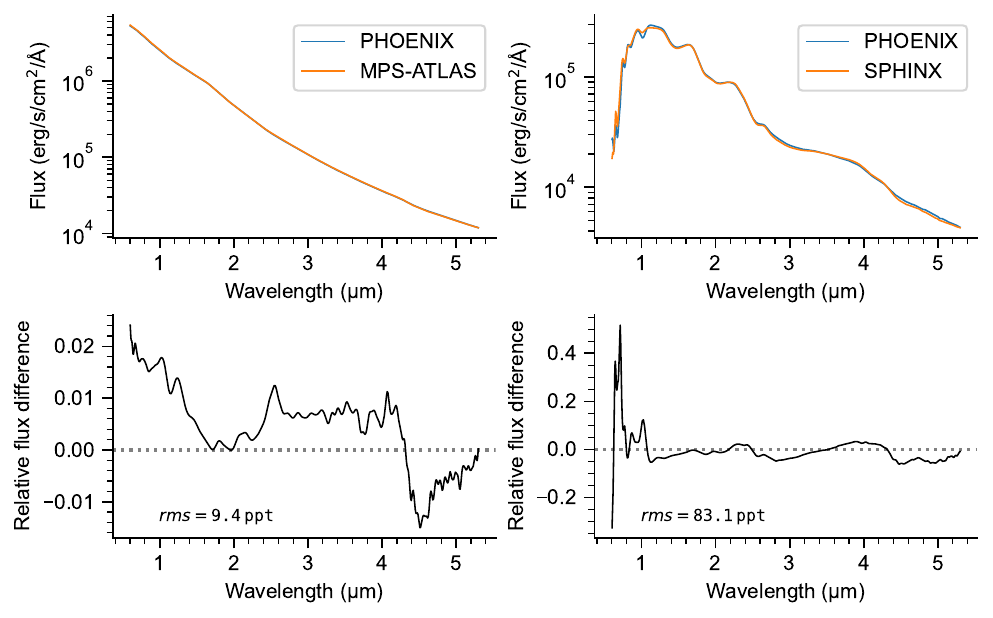}
    \caption{
    State-of-the-art model spectra of quiescent K0 and M6 dwarfs at the wavelengths and resolution of NIRSpec/PRISM.
    The left column shows spectra of a K0 dwarf drawn from the PHOENIX and MPS-ATLAS grids.
    The top panel shows the spectra, and the bottom panel shows the flux difference between them, normalized to the flux of the PHOENIX model.
    The right column shows the same for the PHOENIX and SPHINX spectra of the M6 dwarf used as an example in this work.
    See also \citet[][Figure~1]{Iyer2023} for a comparison of M dwarf spectra across many stellar models.
    \label{fig:stellar_models}
    }
\end{figure*}

We 
explore the impact of imperfections in the stellar spectral models on our inferences
\R{by performing two types of retrievals.
In what we call ``direct'' retrievals, the same stellar spectral model grid is used to simulate the dataset and retrieve on it.
In ``cross''-retrievals, different model grids are used for the data simulation and retrieval.
}
In both cases, the synthetic data are generated using the PHOENIX stellar spectral model grid\footnote{\url{http://phoenix.astro.physik.uni-goettingen.de/}} \citep{Husser2013}.
Relevant to our purposes, the PHOENIX grid spans effective temperatures of $T_\mathrm{eff} \in [2300, 7000]$\,K in 100\,K steps and surface gravities of $\log g \in [0.0, 6.0]$ in steps of 0.5.
For all spectral models, we linearly interpolate between grid points in terms of $T_\mathrm{eff}$, $\mathrm{[Fe/H]}$, and $\log g$ using the \texttt{speclib} package\footnote{\url{https://github.com/brackham/speclib}} \R{\citep{speclib-0.0-beta.0}}.

For the cross-retrievals, we use other model grids to retrieve on the data.
This allows us to examine potential limitations introduced by the models under the assumption that the differences between state-of-the-art models provide a proxy of the differences between the models and reality.
At the sampling of our simulated datasets (see \autoref{sec:precisions}), these differences average ${\sim}10$\,ppt for K0 stars and earlier types and ${\sim}200$\,ppt for M6 stars---with local differences above 100\% (\autoref{fig:stellar_models}).
Considering that planetary signals within reach of \JWST{}'s precision can be of the order of a few hundred parts per million \citep[e.g.,][]{Rustamkulov2023_ERS}, it is crucial to explore how uncertainties stemming from model fidelity will challenge our retrievals that incorporate TLS signals.

\R{Given the} temperature regimes of the state-of-the-art model grids, we used different models for the K0 and M6 cross-retrievals.
For the K0 case, we used the MPS-ATLAS model grid \citep{Witzke2021, Kostogryz2023}\R{, which} spans effective temperatures of $T_\mathrm{eff} \in [3500, 9000]$\,K in 100\,K steps and surface gravities of $\log g \in \{3.0, 3.5, 4.0, 4.2, 4.3, 4.4, 4.5, 4.6, 4.7, 5.0\}$.
For the M6 case, we used the SPHINX model grid \citep{Iyer2023}\R{, which} spans effective temperatures of $T_\mathrm{eff} \in [2000, 4000]$\,K in 100\,K steps and surface gravities of $\log g \in [4.0, 5.5]$ in steps of 0.25.
As with the direct retrievals, we fixed the metallicity ([Fe/H] or [M/H]) of all spectra to 0.
For the SPHINX model grid, we also fixed $\mathrm{C/O} = 0.5$.

\R{The} wavelength range of \R{all} models \R{encompasses} the 0.6--5.3\,$\micron$ range of NIRSpec/PRISM\R{, the instrument for our simulated observations (see \autoref{sec:precisions})}.
\R{However, all} three model grids---PHOENIX, MPS-ATLAS, and SPHINX---are calculated at higher spectral resolutions \R{($R{\sim}100{,}000$--$500{,}000$, $R{\sim}500{,}000$, and $R{\sim}250$, respectively)}
than provided by NIRSpec/PRISM ($R{\sim}100$).
\R{When comparing the models to the data, we convolve the model spectra with a Gaussian kernel to match the resolution of the data and downsample them to match the pixel sampling of the simulated data.}

\subsection{Simulated precisions} \label{sec:precisions}


We simulated the precision of \JWST{} observations of our synthetic targets using PandExo\footnote{\R{\url{https://exoctk.stsci.edu/pandexo/}}} \citep[\R{version 2.0;}][]{PandExo}.
We focus on observations with the Near Infrared Spectrograph (NIRSpec) with the low-resolution ($R{\sim}100$) PRISM disperser, following the approach of the observations of WASP-39b \citep{JTECERST2023, Rustamkulov2023_ERS} through the \JWST{} Transiting Exoplanet Community Early Release Science Program \citep{Bean2018}.

We used NIRSpec in Bright Object Time Series mode with the $1\farcs6 \times 1\farcs6$ fixed-slit aperture (s1600a1) and PRISM disperser.
This setup provides spectra spanning 0.6--5.3\,$\micron$ at a spectral resolving power of $R{\sim}100$.
We also used the SUB512 subarray, five groups per integration, and the NRSRAPID read mode.
We set the total observing time to three times the transit duration.
We assumed a constant noise floor of 10\,ppm, consistent with the $3\sigma$ upper limit of 14\,ppm measured in lab time series \citep{Rustamkulov2022} and the lack of significant systematic errors noted in the observations of WASP-39b \citep{Rustamkulov2023_ERS}.

We note that our choice of $J{=}11$ apparent magnitudes for the host stars places them among the best targets that can be observed with NIRSpec/PRISM, making our simulated datasets among the best single-visit datasets possible with this observing mode.
Combining the data from the full out-of-transit baseline, our simulated spectra have a typical per-pixel signal-to-noise ratio (S/N) of 16,000 (${\sim}$62\,ppm error).
\R{To guard against our results being driven by a single noise instance, we generated five datasets for each stellar spectrum in which we perturb the model truths by noise instances randomly drawn from Gaussian distributions given by the per-pixel uncertainties.}

\section{Retrievals} \label{sec:retrievals}

We retrieved on each of the \R{50} simulated out-of-transit stellar spectra (2 host stars $\times$ 5 activity levels \R{$\times$ 5 noise instances}) using four models, accounting for one, two, three, and four spectral components, respectively.
We refer to these as the ``1-comp,'' ``2-comp,'' ``3-comp,'' and ``4-comp'' models hereafter.
The rationale behind testing this range of model complexity is that it encompasses all of the true complexity of our input models and more, thereby allowing us to assess when biases emerge from our inability to robustly constrain the true complexity of the observed photosphere.
The following provides further details on the models and the retrieval procedure.

\subsection{Model Definition} \label{sec:model}

We model the flux at wavelength $\lambda$ received from each host star $F_\lambda$ as
\begin{equation}
    F_\lambda = \sum_{i=1}^{N} f_i S_{i, \lambda} \left( \frac{R_s}{D_s} \right)^2,
    \label{eq:model}
\end{equation}
in which $f_i$ and $S_{i, \lambda}$ are the filling factors and emergent spectra of the $i$th spectral component present on the stellar disk, $N$ is the number of spectral components, $R_s$ is the stellar radius, and $D_s$ is the stellar distance.
The units of our model and simulated data are erg\,s$^{-1}$\,cm$^{-2}$\,\AA$^{-1}$.

The goal of the retrieval procedure is to identify the values that maximize the likelihood $\mathcal{L}$ of the model ($F_\mathrm{model}$) when compared to the data ($F_\mathrm{data}$).
For the natural logarithm of the likelihood function, we adopt
\begin{equation}
    \ln \mathcal{L} = 
    - \frac{1}{2} \sum 
       \left( \frac{(F_\mathrm{data, \lambda} - F_\mathrm{model, \lambda})^2} 
                   {\sigma_{\lambda}^2} 
            + \ln \left(2 \pi \sigma_{\lambda}^2\right) \right).
\end{equation}
Following \citet{emcee, emcee_v3}\footnote{See \url{https://emcee.readthedocs.io/en/stable/tutorials/line/}, for an example implementation.}, we model the $\sigma_\lambda$ as the quadrature sum of photon noise $\sigma_\mathrm{phot, \lambda}$, given by the simulations in \autoref{sec:precisions}, and an additional noise term $\sigma_\mathrm{jitter, \lambda}$, which encapsulates any additional noise present in the data.
We parameterize the additional noise as a fractional underestimation of the variance following
\begin{equation}
    \sigma_\mathrm{jitter, \lambda} = f_\mathrm{var} F_\mathrm{model, \lambda},
    \label{eq:sigma_jitter}
\end{equation}
which means that the amplitude of $\sigma_\mathrm{jitter, \lambda}$ scales with the model flux.
While we did not inject systematic noise into the simulations, this approach adds a level of realism to our retrievals, effectively inflating the data uncertainty to account for any shortcomings of \R{the} models in describing the data.

\subsection{Priors} \label{sec:priors}

\begin{deluxetable}{lll}[t]
\tablecaption{Free Parameters and Their Priors for the Four Retrieval Models. \label{tab:models}}
\tablehead{Parameter & Prior & Unit}
\startdata
$T_{n \in \{1, 2, 3, 4\}}$ & \U{2300}{5500} & K \\
$f_{n \in \{1, 2, 3, 4\}}$ & \D{0}{1}       & ... \\
$R_s$                      & \U{0.08}{1.00} & R$_\sun$ \\
$\ln f_\mathrm{var}$      & \U{-50}{0}     & ... \\
\enddata
\tablecomments{
    \U{a}{b} designates a uniform prior over the range $(a, b)$.
    \R{\D{a}{b} designates a Dirichlet prior over the range $(a, b)$.}
}
\end{deluxetable}

\autoref{tab:models} summarizes the free parameters for our four models and their priors.
We parameterize the spectral components by their temperatures ($T_1$, $T_2$ $T_3$, and $T_4$), and we place wide, uniform priors on all temperatures.
\R{We place a joint Dirichlet prior\footnote{See \url{https://johannesbuchner.github.io/UltraNest/priors.html}.} on all filling factors ($f_1$, $f_2$, $f_3$, and $f_4$) such that their values sum to 1.}
To prevent degenerate solutions and automatically ensure that the number order of the components corresponds to their prevalence on the stellar disk, 
\R{we enforce $f_1 > f_2 > f_3 > f_4$ with a likelihood penalty.}

The stellar radius and distance are fully degenerate parameters in our model (\autoref{eq:model}).
Rather than assuming errors to use for normal priors on these parameters, we fix $D_s$ to the adopted distance and fit for $R_s$ with a uniform prior.
We also fix the stellar metallicity [Fe/H] to 0 and the surface gravity $\log g$ to the value given by the mass and radius of host star provided in \autoref{sec:system_parameters}.

The final free parameter in each model is the fractional underestimation of the variance $f_\mathrm{var}$, which accounts for additional noise in the data (\autoref{eq:sigma_jitter}).
To ensure that it is always positive, and to allow the sampling to explore a large dynamic range, we fit for \R{$\ln f_\mathrm{var}$} with a uniform prior of $(-50, 0)$.
We note that the median value of \R{$\ln (\sigma_\mathrm{phot, \lambda} / F_\mathrm{data, \lambda})$} is $-10$, and so the parameter space included in this prior spans scenarios in which systematic errors are many orders of magnitude below or above the photon noise.

In total, there are 3, \R{6, 8, and 10} free parameters for the 1-comp, 2-comp, 3-comp, and 4-comp models, respectively.

\subsection{Model Inference} \label{sec:model_inference}

We derive the posterior probability distributions of the model parameters with the nested sampling Monte Carlo algorithm MLFriends \citep{Buchner2014, Buchner2017} using the \texttt{UltraNest}\footnote{\url{https://johannesbuchner.github.io/UltraNest/}} Python package \citep{Buchner2021}.
We use slice sampling to efficiently explore the parameter space, \R{using 400 live points and} defining the number of steps as 10 times the number of parameters.
At each sampling step, we use the \texttt{speclib}\footnote{\url{https://github.com/brackham/speclib}} Python package \R{\citep{speclib-0.0-beta.0}} to generate the component spectra included in the model.
We use the \texttt{SpectralGrid} object within \texttt{speclib} to do this efficiently, loading a spectral grid into memory once with the fixed metallicity and surface gravity values and linearly interpolating between temperature grid points to produce the sample spectra.
We note that linear interpolation is likely not the best approach in the high-S/N regime in which we are operating here.
We discuss this complication further in \autoref{sec:discussion}.

\subsection{Model Selection} \label{sec:model_selection}

Our studied parameter space covers \R{50} simulated datasets (2 host stars $\times$ 5 activity levels \R{$\times$ 5 noise instances}).
We retrieve on each using two spectral grids, the PHOENIX grid \R{for ``direct'' retrievals} or another \R{for ``cross''-retrievals (MPS-ATLAS for the K dwarf and SPHINX for the M dwarf)}.
For each dataset--grid pair, we would like to test our four model complexities (1-comp to 4-comp) and determine which model best describes the data.

\R{We approach model selection in two ways. In the first, we use the Bayesian evidence ($\ln \mathcal{Z}$) of each model computed by \texttt{UltraNest}} as the basis for model selection.
We define the best model as the simplest model that produces a significantly better fit than other models.
We adopt a Bayes factor of $\Delta \ln \mathcal{Z} = 5.0$ as the threshold for significance, corresponding to an odds ratio of ${\sim}150:1$ \citep{Trotta2008} or a $3.6\sigma$ result \citep{Benneke2013}.
In other words, we selected a more complex model over a simpler one only when it provides a marginal increase in the $\ln \mathcal{Z}$ of 5.0 or more.

\R{
In the second approach, we use the Akaike information criterion \citep[AIC;][]{Akaike1974} for model selection.
Given the AIC values for the four model complexities for a given dataset--grid pair, we calculate the Akaike weights for each model.
We then define the best model as that which gives the highest Akaike weight.
We note that in our case the AIC modified for small sample sizes (AICc) is nearly identical to the AIC because the number of data points is much larger than the number of fitted parameters.
}

In total, the \R{400} nested sampling retrievals in this analysis cover two host stars, five activity levels, two spectral model grids, four model complexities\R{, and five noise realizations}.


\begin{figure*}[htp]
    \centering
    \includegraphics[width=\textwidth]{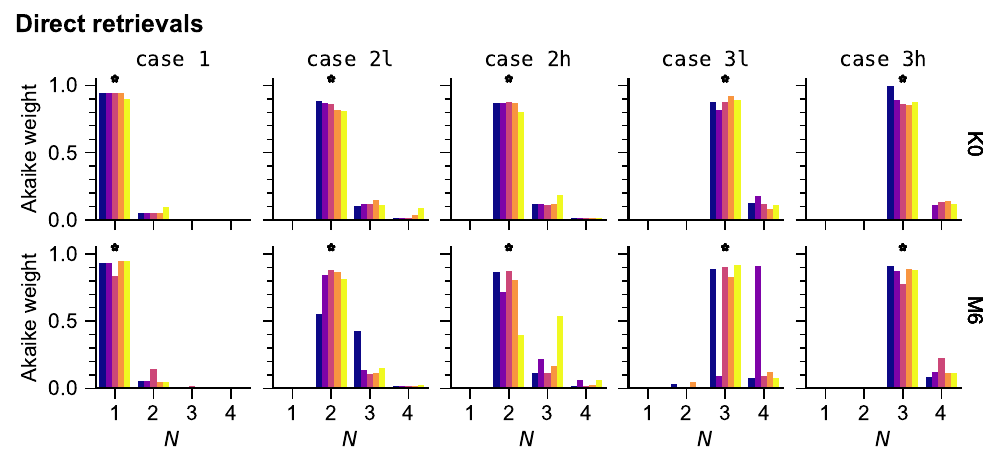}
    \caption{
    \R{Akaike weights} for model fits with different levels of complexity in the case of the direct retrievals.
    Here the same model grid was used to simulate and retrieve on the data.
    The top \R{and bottom rows give} results for the K0 \R{and M6 stars, respectively}.
    From left to right, the \R{panels} give results for the \texttt{1}, \texttt{2l}, \texttt{2h}, \texttt{3l}, and \texttt{3h} cases, respectively.
    In each panel, the \R{Akaike weights are shown for the models with $N$ spectral components, and the correct model complexity is indicated by a star.}
    \R{The colors distinguish results for the five noise models.}
    \R{For the direct retrievals, the Akaike weights generally favor the correct model complexity, though the noise instance can change this, and other models sometimes receive substantial weight.}
    \label{fig:direct_retrievals_AIC_weights}
    }
\end{figure*}

\section{Results} \label{sec:results}

In analyzing the results of the retrievals, we focus on \R{four} topics: whether we can infer the correct level of complexity, whether we can retrieve the correct \R{photospheric parameters, whether we can then calculate the proper correction and thus reduce biases on the transmission spectrum}, and the impact of accounting for the heterogeneity on the uncertainty budget.
We present each of these topics in turn in the following.

\subsection{Inferring the Correct Level of Complexity} \label{sec:complexity}

\R{\subsubsection{Direct Retrievals}}


\begin{figure*}[htp]
    \centering
    \includegraphics[width=\textwidth]{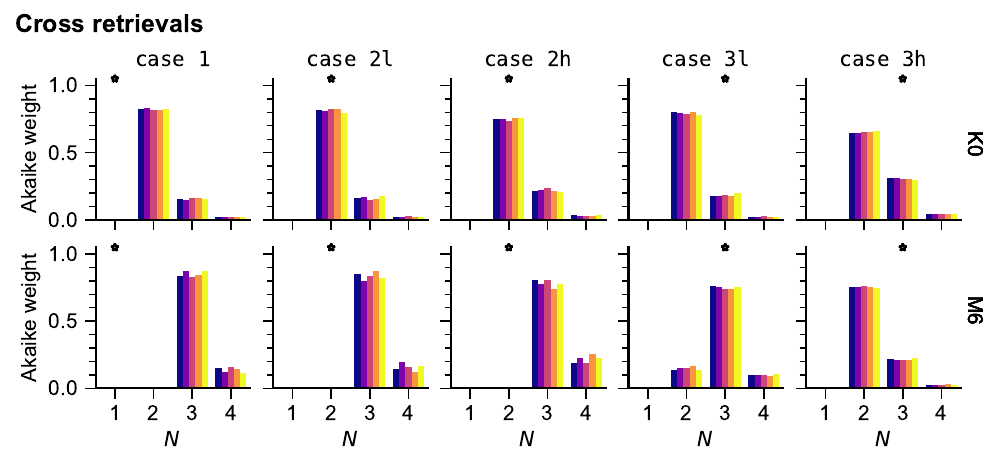}
    \caption{
    \R{Akaike weights} for model fits with different levels of complexity in the case of the cross-retrievals.
    The figure elements are the same as those in \autoref{fig:direct_retrievals_AIC_weights}.
    The top row gives results for the K0 star, simulated with PHOENIX and retrieved with MPS-ATLAS.
    The bottom row gives results for the M6 star, simulated with PHOENIX and retrieved with SPHINX.
    In most cases, the cross-retrievals (i.e., \R{retrievals with lower} model fidelity) fail to \R{weight models with} the correct complexity.
    \label{fig:cross_retrievals_AIC_weights}
    }
\end{figure*}

We \R{first review} the results for the direct-retrieval cases, which assume model fidelity.
As detailed in \autoref{sec:model_selection}, we analyzed \R{50} simulated spectra \R{(10 input spectra $\times$ 5 noise realizations)}, trying to fit to each four models with varying levels of complexity.
\R{
In terms of the Bayesian-evidence-based criteria laid out in \autoref{sec:model_selection}, we find that the correct level of complexity was inferred in 46 out of 50 cases.
The four exceptions were all different noise realizations of the M6 \texttt{case 3l} dataset, for which the 2-comp model was preferred.
In terms of the Akaike-weight-based criteria (\autoref{sec:model_selection}), the correct level of complexity was inferred in 48 out of 50 cases (\autoref{fig:direct_retrievals_AIC_weights}).
The exceptions were single noise instances of the M6 \texttt{case 2l} and \texttt{3l} datasets.
Interestingly, in the former dataset, the correct model complexity still receives a substantial Akaike weight.
This motivates us to explore corrections based on Bayesian model averaging using the Akaike weights, which we discuss in \autoref{sec:corrections}.
Further goodness-of-fit details are given in \autoref{appendix:gof}.
}

\R{\subsubsection{Cross-retrievals}}
\label{sec:cross_results}

We now turn to the results of the cross-retrieval cases.
\R{Here the results are similar for all noise instances.
In terms of Bayesian evidence, the correct complexity was inferred in 15 out of 50 cases.
Regardless of noise instance, the correct complexity was selected for the K0 \texttt{case 2l} and \texttt{2h} datasets and M6 \texttt{case 2l} datasets.
In terms of Akaike weight, the results are also similar for all noise instances (\autoref{fig:cross_retrievals_AIC_weights}).
The 2-comp models were most highly weighted for all K0 spectra, and the 3-comp models were most highly weighted for all M6 spectra except the M6 \texttt{case 3h} spectra, for which the 2-comp models were favored.
As a result, based on the Akaike weights, the correct model complexity was selected in only 15 out of 50 cases.
}


Nonetheless, whether the cross-retrievals happened to identify the right level of complexity or not, they universally provide poor fits to the data \R{in terms of reduced $\chi^2$ ($\chi^2_{r} \gg 1$ before inflating errors)} in this high-S/N regime.
\R{While} PHOENIX model fits have \R{$\chi^2_r \approx 1$} before accounting for the inflated uncertainties \R{(\autoref{tab:gof_results_direct})}, the corresponding values for the cross-retrievals are $\chi^2_r \sim 10^4$ for the MPS-ATLAS models and \R{$\chi^2_r \sim 10^5$} for the SPHINX models \R{(\autoref{tab:gof_results_cross})}.
\R{This underscores that intermodel differences are substantially larger than the measurement precision of the synthetic spectra used.}

\begin{figure}
    \centering
    \includegraphics[width=\columnwidth]{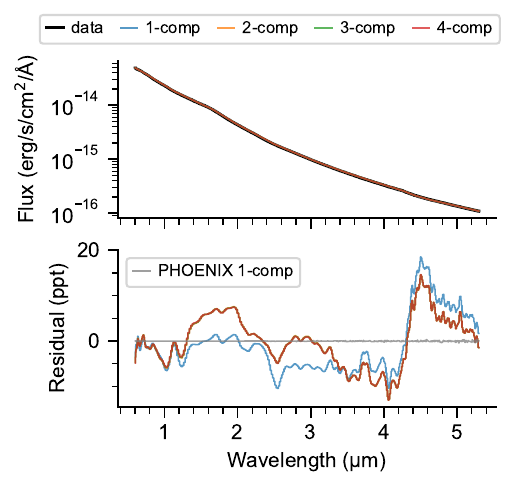}
    \caption{
    Best fits to the K0 \texttt{case 1} spectrum for MPS-ATLAS models with one to four components.
    The top panel shows the simulated NIRSpec/PRISM spectrum in black, along with the four fits.
    Uncertainties for the data and posterior models are smaller than the line widths.
    Likewise, differences between the data, simulated with the PHOENIX grid, and the models, interpolated from the MPS-ATLAS grid, are too small to be apparent at this scale.
    The bottom panel shows the residuals for the fits, normalized to the data.
    In this view, residuals on the order of 10\,ppt (1\%) are evident.
    For comparison, the gray line shows the residuals for a direct retrieval with the best-fit one-component PHOENIX model.
    The relative scales of the residuals for the cross- and direct retrievals gives a sense of the impact of uncertainties due to stellar models vs. photon noise.
    \label{fig:k0_case1_fit}
    }
\end{figure}

As an example, we highlight the K0 \texttt{case 1} cross-retrieval with the MPS-ATLAS grid (\autoref{fig:k0_case1_fit}).
Like our other simulated spectra, this spectrum has a typical per-pixel S/N of 16,000 (${\sim}$62\,ppm error).
At this precision, the differences between the PHOENIX spectra used to simulate the data and the MPS-ATLAS spectra used in the retrieval are readily apparent.
The bottom panel of \autoref{fig:k0_case1_fit} shows that the residuals for all cross-retrieval models are many orders of magnitude higher than those of the correctly inferred direct-retrieval model.
This example underscores that the fidelity of the model grid\R{, i.e., its ability to accurately describe the data,} is crucially important for arriving at the appropriate inferences.

We caution that this does not mean that one should simply select the model grid that provides the best fits when inferring photospheric properties from out-of-transit spectra.
Instead, the results of this exercise raise concerns about model-based inferences of photospheric heterogeneity in general, assuming that the differences between modern model spectra provide a proxy for the differences between models and actual spectra of photospheric components.
We return to this point in the discussion.

\R{\subsection{Inferring Stellar Parameters} \label{sec:parameter_inference}}

\begin{deluxetable*}{cccCCCCCCCCCCCCCCCCCC}
\tablecaption{
    \R{Stellar Parameter Uncertainties and Biases for Preferred Models.}
    \label{tab:photosphere_params}
}
\tabletypesize{\scriptsize}
\tablehead{
    \colhead{Grid} & \colhead{Star} & \colhead{Case} & \colhead{Model} & 
    \multicolumn{2}{c}{$R_s$} & \multicolumn{2}{c}{$T_1$} & \multicolumn{2}{c}{$T_2$} & \multicolumn{2}{c}{$T_3$} & 
    \multicolumn{2}{c}{$f_1$} & \multicolumn{2}{c}{$f_2$} & \multicolumn{2}{c}{$f_3$} \\
    \colhead{} & \colhead{} & \colhead{} & \colhead{} &
    \colhead{$\sigma\,(R_\sun)$} & \colhead{$z$} & 
    \colhead{$\sigma$\,(K)} & \colhead{$z$} & 
    \colhead{$\sigma$\,(K)} & \colhead{$z$} & 
    \colhead{$\sigma$\,(K)} & \colhead{$z$} & 
    \colhead{$\sigma$} & \colhead{$z$} & 
    \colhead{$\sigma$} & \colhead{$z$} & 
    \colhead{$\sigma$} & \colhead{$z$}
}
\startdata
P & K0 & $\texttt{1}$ & 1 & 1.5\times10^{-6}  & -1.3 & 0.0079 & 0.80 &  &  &  &  &  &  &  &  &  &  \\
P & K0 & $\texttt{2l}$ & 2 & 5.8\times10^{-6} & 0.13 & 0.063 & 0.43 & 4.9 & 0.72 &  &  & 8.0\times10^{-5} & -0.62 & 8.0\times10^{-5} & 0.62 &  &  \\
P & K0 & $\texttt{2h}$ & 2 & 6.0\times10^{-6} & -0.44 & 0.068 & -0.45 & 0.97 & -0.065 &  &  & 8.3\times10^{-5} & 0.40 & 8.3\times10^{-5} & -0.40 &  &  \\
P & K0 & $\texttt{3l}$ & 3 & 9.7\times10^{-6} & -0.31 & 19 & -1.8 & 35 & \mathbf{4.7} & 13 & 0.34 & 0.063 & \mathbf{3.2} & 0.064 & \mathbf{-3.2} & 0.00020 & 0.20 \\
P & K0 & $\texttt{3h}$ & 3 & 9.6\times10^{-6} & -2.1 & 4.7 & 1.9 & 8.7 & 0.44 & 2.5 & 2.1 & 0.048 & 1.2 & 0.048 & -1.2 & 0.00018 & 2.1 \\
P & M6 & $\texttt{1}$ & 1 & 7.7\times10^{-7}  & -0.13 & 0.010 & 0.10 &  &  &  &  &  &  &  &  &  &  \\
P & M6 & $\texttt{2l}$ & 2 & 1.8\times10^{-6} & -0.46 & 0.043 & 0.56 & 8.0 & 1.2 &  &  & 0.00025 & -0.95 & 0.00025 & 0.95 &  &  \\
P & M6 & $\texttt{2h}$ & 2 & 1.8\times10^{-6} & 0.023 & 0.045 & 0.13 & 1.5 & 0.22 &  &  & 0.00025 & -0.29 & 0.00025 & 0.29 &  &  \\
P & M6 & $\texttt{3l}$ & 3 & 1.8\times10^{-6} & 1.9 & 3.3 & -1.6 & 38 & 0.54 & 19 & -1.6 & 0.045 & -0.29 & 0.045 & 0.34 & 0.0016 & -1.2 \\
P & M6 & $\texttt{3h}$ & 3 & 1.9\times10^{-6} & 1.5 & 4.2 & \mathbf{-5.8} & 250 & 0.87 & 300 & -0.95 & 0.05 & \mathbf{-4.5} & 0.034 & \mathbf{6.3} & 0.018 & 0.82 \\
M & K0 & $\texttt{1}$ & 1 & 0.00028 & \mathbf{-7.5} & 1.9 & \mathbf{15} &  &  &  &  &  &  &  &  &  &  \\
M & K0 & $\texttt{2l}$ & 2 & 0.00061 & \mathbf{-14} & 5.5 & -2.4 & 55 & \mathbf{4.7} &  &  & 0.0057 & \mathbf{9.4} & 0.0057 & \mathbf{-9.4} &  &  \\
M & K0 & $\texttt{2h}$ & 2 & 0.00062 & \mathbf{-15} & 7.5 & -0.99 & 55 & \mathbf{4.5} &  &  & 0.0074 & \mathbf{6.2} & 0.0074 & \mathbf{-6.2} &  &  \\
M & K0 & $\texttt{3l}$ & 3 & 0.00056 & \mathbf{-15} & 25 & -0.58 & 780 & 0.68 & 570 & 0.010 & 0.13 & 1.0 & 0.12 & -0.78 & 0.020 & -2.1 \\
M & K0 & $\texttt{3h}$ & 3 & 0.00058 & \mathbf{-15} & 28 & -0.5 & 840 & 0.66 & 700 & -0.21 & 0.14 & -0.46 & 0.12 & 0.75 & 0.031 & -0.72 \\
S & M6 & $\texttt{1}$ & 1 & 0.00041 & -1.4 & 3.3 & 0.53 &  &  &  &  &  &  &  &  &  &  \\
S & M6 & $\texttt{2l}$ & 2 & 0.00032 & \mathbf{-9.8} & 2.5 & \mathbf{19} & 87 & \mathbf{-17} &  &  & 0.0016 & -0.87 & 0.0016 & 0.87 &  &  \\
S & M6 & $\texttt{2h}$ & 2 & 0.00030 & \mathbf{-6.9} & 2.4 & \mathbf{20} & 88 & \mathbf{-17} &  &  & 0.0054 & \mathbf{-7.8} & 0.0054 & \mathbf{7.8} &  &  \\
S & M6 & $\texttt{3l}$ & 3 & 0.00074 & \mathbf{-7.1} & 12 & \mathbf{3.1} & 39 & \mathbf{6.0} & 77 & \mathbf{-19} & 0.11 & 2.3 & 0.11 & -2.3 & 0.0023 & -1.4 \\
S & M6 & $\texttt{3h}$ & 3 & 0.00071 & \mathbf{-6.1} & 13 & 2.9 & 370 & 0.25 & 370 & -3.8 & 0.15 & -0.56 & 0.15 & 0.33 & 0.0038 & \mathbf{9.2}
\enddata
\tablecomments{
    Results are shown for the first noise instance.
    The grids are PHOENIX (P), MPS-ATLAS (M), and SPHINX (S).
    Uncertainties are given in terms of the standard deviation of the posterior sample ($\sigma$).
    Biases are given in terms of $z$-score ($z$) for the true value relative to the mean and standard deviation of the posterior sample, with $|z| > 3$ in bold.
    For clarity, means of the posterior samples are not shown.
}
\end{deluxetable*}

\R{
\autoref{tab:photosphere_params} summarizes the uncertainties and biases on the retrieved stellar parameters.
The results for the models reflecting the true complexity and in the case of the first noise instance are included.
For the direct retrievals, the median standard deviations on all inferred temperatures and covering fractions are 4.5\,K and 930\,ppm, respectively.
The mean of the posterior sample is within $1\sigma$ of the true value for most parameters.
}

\R{
Only six parameters, all from the K0 \texttt{case 3l} and M6 \texttt{case 3h} fits, are more than $3\sigma$ from truth.
The parameter uncertainties for these two fits are also much larger than for other direct retrievals.
Closer inspection of the results reveals that the posterior distributions are multimodal, meaning that the nested sampling algorithm identified other possible parameter combinations that provide fits comparable to that of the input parameters.
Given the prevalence of these other modes, the means of the posterior parameter distributions are thus biased with respect to the true values.
Turning to the results from the other noise instances, we find that this behavior is dependent on the noise instance for the K6 \texttt{case 3l} retrieval, highlighting the role that even small levels of noise can play in this sort of analysis.
On the other hand, the results for the M6 \texttt{case 3h} spectrum are similar for all noise instances, which suggests an inherent limitation in this sort of analysis for highly active M dwarfs.
}

\R{
For the cross-retrievals, the precision and accuracy of all posterior inferences are notably worse.
The median standard deviations on all inferred temperatures and covering fractions are 47\,K and 1.4\% (14,000\,ppm), respectively.
The mean of the posterior sample is within $5\sigma$ of the true value for only 29 of the 52 inferred parameters.
In short, while the stellar parameters are generally estimated accurately and precisely for the direct retrievals, estimates from the cross-retrievals are both biased and relatively imprecise.
}

\subsection{Inferring Corrections and Reducing Biases} \label{sec:corrections}

\begin{figure*}
    \centering
    \includegraphics[width=\textwidth]{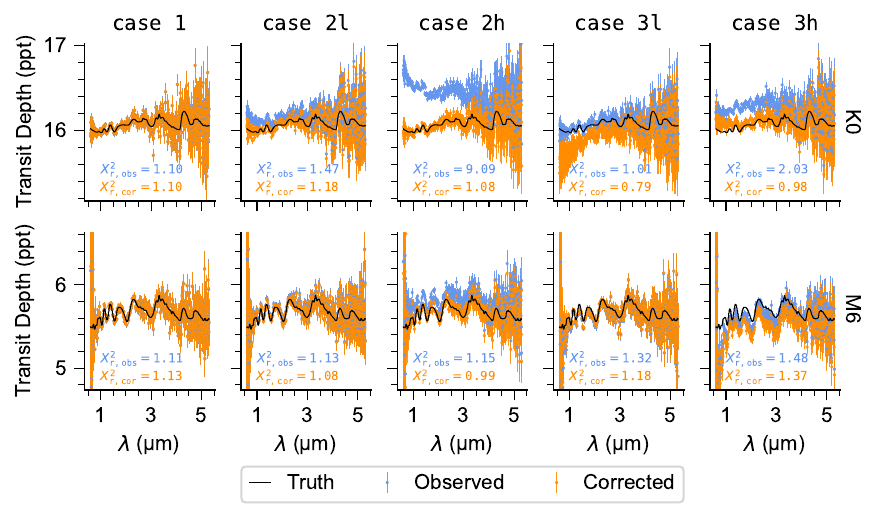}
    \caption{Change in bias of transmission spectra after applying corrections implied by the direct retrievals.
    The top and bottom panels show results for the K0 and M6 stars, respectively.
    From left to right, the columns correspond to $\texttt{cases 1}$, $\texttt{2l}$, $\texttt{2h}$, $\texttt{3l}$, and $\texttt{3h}$.
    In each panel the black lines give the true transmission spectra of the planet (warm Jupiter in the top row, super-Earth in the bottom), and the blue and orange points show the observed \R{(}contaminated\R{)} and corrected transmission spectrum, respectively.
    The \R{reduced $\chi^2$ values ($\chi^2_r$)} of the residuals \R{with respect to the true transmission spectrum for} the observed and corrected spectra are shown in blue and orange, respectively.
    \R{Uniform offsets (not shown) are applied to minimize $\chi^2_r$ in all cases.}
    \R{The corrections reduce the bias for the cases with two or three types of photospheric heterogeneities, either by bringing the transmission spectra more in line with the true values (e.g., $\texttt{cases 2l}$ and $\texttt{2h}$) or by increasing the uncertainties to reflect our imperfect understanding of the stellar photospheric properties (e.g., $\texttt{cases 3l}$ and $\texttt{3h}$).}
    \label{fig:bias_reduction}
    }
\end{figure*}

\R{
We are now interested in whether our estimates of the stellar parameters enable stellar contamination corrections that can reduce bias in the transmission spectra.
In light of the cross-retrievals' overall poor fits (\autoref{sec:cross_results}) and degraded inferences (\autoref{sec:parameter_inference}), we} now focus on the direct retrievals only.
\R{
While the AIC generally favored} the correct level of complexity \R{for these retrievals (\autoref{sec:complexity}), we use Bayesian model averaging here to incorporate properly weighted information from fits of all model complexities in the correction calculation.
The results from all noise instances are similar, and so for simplicity we again focus on the first noise instance.
}

\R{\subsubsection{Calculating True and Inferred Stellar Contamination}}

We calculate the impact of the photospheric heterogeneity on the transmission spectrum as
\begin{equation}
    \epsilon_\lambda = \frac{S_{1, \lambda}}
                    {\sum_{i=1}^N f_i S_{i, \lambda}}
    \label{eq:epsilon}
\end{equation}
in which $S_i$ is the spectrum of the $i$th spectral component and $f_i$ is its filling factor.
This expression is equivalent to those presented by \citet{Rackham2018, Rackham2019} but terms are rearranged to clearly convey their origin.
The numerator corresponds to the mean stellar spectrum illuminating the exoplanet atmosphere during the transit, whereas the denominator corresponds to the full-disk stellar spectrum observed outside of the transit.
The observed transmission spectrum is then
\begin{equation}
    D_\mathrm{obs, \lambda} = \epsilon_\lambda D_\mathrm{true, \lambda},
\label{eq:D}
\end{equation}
in which $D_\mathrm{true, \lambda}$ is the true planetary transmission spectrum, i.e., the square of the wavelength-dependent planet-to-star radius ratio $(R_{p,\lambda}/R_s)^2$.
When $N = 1$, $\epsilon_\lambda = 1$ and there is no contamination.
The implicit assumption with \autoref{eq:epsilon} is that the planet transits the dominant spectral component, whereas the other spectral components are present elsewhere on the stellar disk.
This is due to our focus in this study on indirectly constraining heterogeneities whose presence cannot be inferred directly through occultations by the transiting exoplanet \R{\citep[e.g.,][]{Espinoza2019, Fu2022}}.
In the case where multiple spectral components are present in the transit chord, the numerator of \autoref{eq:epsilon} can be replaced with another summation using the filling factors of the components within the transit chord \R{\citep[][Equation~7]{ZhangZhanbo2018}}.

We calculated the posterior samples of $\epsilon_\lambda$ using the parameter values at each step in the \R{equally weighted posterior sample}.
\R{We used the Akaike weights shown in \autoref{fig:direct_retrievals_AIC_weights} to properly weight the estimates from each model complexity and create a final, model-averaged posterior sample of $\epsilon_\lambda$.}
We then calculated the inferred values of $D_\mathrm{true, \lambda}$ using \autoref{eq:D} and propagating the measurement uncertainties of $D_\mathrm{obs, \lambda}$ and $\epsilon_\lambda$ through the equation using a Monte Carlo approach.
\R{For each step in the sampling, this included drawing a sample of $D_\mathrm{obs, \lambda}$ using the means and standard deviations of the data points and dividing it by the sample value of $\epsilon_\lambda$ to build a posterior sample of the corrected transmission spectrum, of which we report the mean and standard deviation.}
To distinguish our inferences from actual true values of the transmission spectra, which we know in this exercise, we refer to our inferences as $D_\mathrm{cor, \lambda}$ hereafter.

\R{\subsubsection{Corrected Transmission Spectra}}

\autoref{fig:bias_reduction} shows the observed and corrected transmission spectra from the direct retrievals using the \R{model-averaged results for the stellar contamination signal}.
\R{As shown previously \citep[e.g.,][]{Rackham2018, Rackham2019}, the impact of the TLS effect is strongest at shorter wavelengths, though in the high-activity cases the heterogeneous photosphere can alter the observed transmission spectrum well into the near-infrared (${\sim}3\,\micron$).}
To assess the change in bias, we calculated the \R{reduced $\chi^2$ ($\chi^2_r$)} between the data and the model truth for both $D_\mathrm{obs, \lambda}$ and $D_\mathrm{cor, \lambda}$.
\R{As transmission spectra are largely insensitive to uniform shifts, we apply achromatic offsets (not shown in \autoref{fig:bias_reduction}) to minimize $\chi^2_r$ in all fits.}

We find that the \R{stellar contamination} corrections reduce \R{$\chi^2_r$ and thus} the bias in the transmission spectra in \R{8 of 10} spectra\R{.} 
\R{
For the two spectra from \texttt{case 1}, stellar contamination is not an issue and $\epsilon_\lambda = 1$ by definition.
In this sense, the fact that the correction only marginally increases $\chi^2_r$ (by 0.004 and 0.021 for the K0 and M6 \texttt{case 1} spectra, respectively) is a good result because, in principle, the Akaike weights could have more strongly favored the 2-comp to 4-comp models, leading us to inject bias in the transmission spectrum with a greater unnecessary stellar contamination correction.
In total, the corrections reduced the $\chi^2_r$
}
between the data and the true planetary transmission spectrum by \R{1.00} on average, with the largest \R{reduction being 8.01 (K0 \texttt{case 2h}) and the worst outcome being the increase of 0.002 for the M6 \texttt{case 1} spectrum.}

\R{
The fits to the K0 \texttt{case 3l} and M6 \texttt{case 3h} spectra warrant special attention.
While the corrections ultimately reduced $\chi^2_r$, \autoref{fig:bias_reduction} shows that, without considering the achromatic offset, the corrected spectra are actually further from the model truth than the observed spectra.
As discussed in \autoref{sec:parameter_inference} and shown in \autoref{tab:photosphere_params}, these are the two fits for which the photospheric parameter inferences are both biased and less precise owing to multimodal posterior distributions.
In effect, \autoref{fig:bias_reduction} shows that the biased and less precise estimates of the photospheric parameters ultimately translate to biased stellar contamination corrections and higher uncertainties on the corrected transmission spectra.
As noted in \autoref{sec:parameter_inference}, this behavior depends on the noise instance for the K6 \texttt{case 3l} retrieval but persists for all noise instances for the M6 \texttt{case 3h} retrieval, suggesting an inherent limitation in relying on model spectra to derive stellar contamination corrections for highly active M dwarfs.
Thus, while relying
} on ``perfect'' spectral models but producing \R{biased} inferences, these results both underscore the limitations of this approach in general in this highly \R{nonlinear} parameter space, a point we return to in the discussion.

\subsection{Impact on the Uncertainty Budget}

The final result that we consider here is the impact on the uncertainty budget.
We are interested in the impact of applying the derived corrections from the selected models for the stellar photosphere and propagating their uncertainties on the final uncertainties of the transmission spectra.
\R{
\autoref{fig:bias_reduction} again illustrates the primary result in this context for the direct retrievals.
It shows that the median per-pixel uncertainty on all transmission spectra increased 7\% from 90 to 96\,ppm.
Thus, we find that, for the direct retrievals, the stellar contamination correction contributes negligibly to the overall uncertainty budget.
This highlights that when the model fidelity is high, we are indeed in a regime where measurement uncertainties dominate and the instruments can be used to their maximum potential.
}

\begin{figure*}
    \centering
    \includegraphics[width=\textwidth]{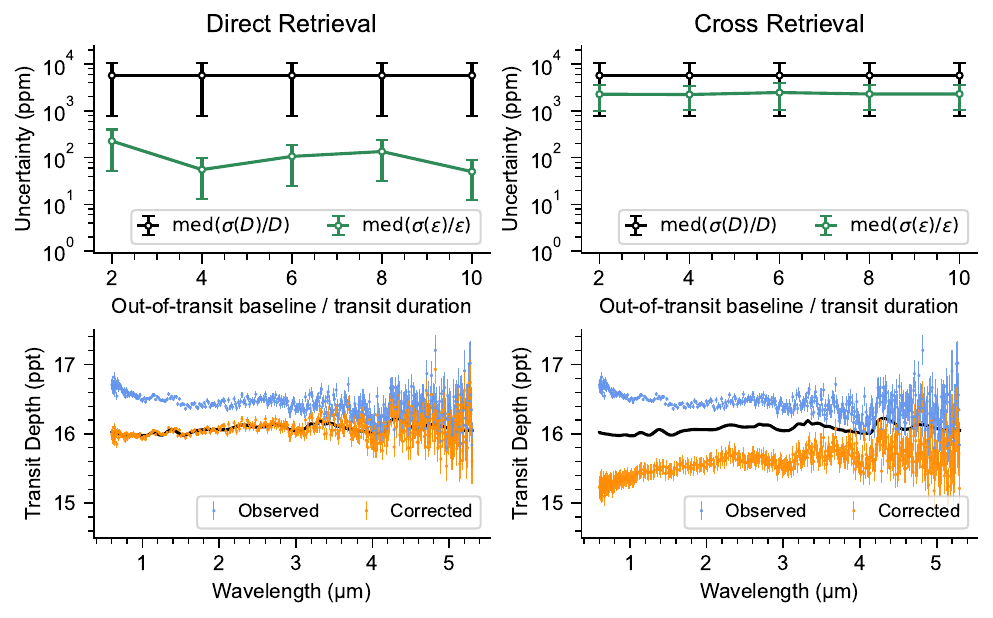}
    \caption{
    Median \R{fractional} uncertainties of the transmission spectrum\R{, $\mathrm{med}(\sigma(D)/D)$,} and the stellar contamination signal\R{, $\mathrm{med}(\sigma(\epsilon)/\epsilon)$,} for the K0 \texttt{case 2h} dataset.
    The left (right) column shows \R{BMA results} for the direct retrieval (cross-retrieval).
    In the top row, the black points show the median fractional uncertainty on the transit depth as a function of the out-of-transit baseline, and the green points show the same for the stellar contamination signal.
    \R{For the direct retrieval, \R{$\mathrm{med}(\sigma(\epsilon)/\epsilon)$}} is consistently smaller \R{than \R{$\mathrm{med}(\sigma(D)/D)$},} and \R{it tends to decrease} with increasing out-of-transit baseline, as expected.
    \R{For the cross-retrieval, $\mathrm{med}(\sigma(\epsilon)/\epsilon)$ is comparable to $\mathrm{med}(\sigma(D)/D)$ and does not decrease with increasing baseline because it is limited by the model fidelity, not the photon noise.}
    The bottom row shows the true (black), observed (blue), and corrected (orange) transmission spectra resulting from these retrievals.
    The \R{direct-retrieval} results show that when the model fidelity is sufficient, \R{an accurate stellar contamination correction is possible and its} contribution to the noise budget of the planetary spectrum is negligible.
    By contrast, the \R{cross-retrieval} results show that \R{with limited model fidelity, $\mathrm{med}(\sigma(\epsilon)/\epsilon)$ is on par with $\mathrm{med}(\sigma(D)/D)$, yielding uncertainties on the corrected transmission spectrum that are roughly twice as large and a stellar contamination correction that is not guaranteed to be accurate.}
    \label{fig:relative_uncertainties}
    }
\end{figure*}

\R{Turning to the cross-retrievals,} \autoref{fig:relative_uncertainties} illustrates the most salient point in this context, which is that the ultimate uncertainty contribution of model-based corrections for stellar contamination depends strongly on how well the models are able to describe the true spectra behind the data (i.e., the ``model fidelity'').
Focusing on the M6 \texttt{case 2h} dataset, this figure shows that in the case of the direct retrieval the \R{median} relative uncertainty on $\epsilon$\R{, $\mathrm{med}(\sigma(\epsilon)/\epsilon) = 230$\,ppm}, is vanishingly small compared to \R{the relative uncertainty on the transit depth, $\mathrm{med}(\sigma(D)/D) = 5700$\,ppm}, due to the high degree of fidelity between the synthetic stellar spectrum and the PHOENIX-based retrieval\R{. This leads} to a correction that imparts no notable additional uncertainty on the final transmission spectrum \R{when propagating uncertainties through \autoref{eq:D}}.
\R{Thus, when aiming for a per-point uncertainty of, say, 50\,ppm on a 2\% transit depth \citep[e.g.,][]{Rustamkulov2023_ERS} and employing stellar models for a stellar contamination correction, the discrepancy between the stellar spectrum and any stellar model used should be less than 50\,ppm / 2\% = 2.5\,ppt in order for the correction to not dominate the final uncertainty.}

On the other hand, in the cross-retrieval case the relative uncertainty on $\epsilon$ is \R{an order} of magnitude larger \R{($\mathrm{med}(\sigma(\epsilon)/\epsilon) = 2300$\,ppm)}, stemming from the need to inflate uncertainties to produce adequate fits \R{($\chi^2_r \approx 1$)}\R{.}
\R{Thus, } the stellar contamination correction \R{contributes appreciably to} the final uncertainty of the transmission spectrum\R{, and the median per-pixel uncertainty of the corrected spectrum is 45\% larger.}
\R{Moreover, due to the lack of model fidelity, the correction is improper, and the ``corrected'' spectrum exhibits as much bias with respect to the true spectrum as the observed spectrum.}

\R{\subsubsection{Impact of the Baseline Duration}}

We also explored how these results depend on the duration of the out-of-transit baseline, repeating our entire retrieval analysis with simulated datasets \R{with longer out-of-transit baselines, from two times the transit duration (our nominal case) to 10 times it (\autoref{fig:relative_uncertainties}, top row)}.
\R{For the direct retrievals, we} find that \R{$\sigma(\epsilon)/\epsilon$ decreases} with increasing baseline, as expected, 
\R{
though it is already much lower than $\sigma(D)/D$ with even the shortest out-of-transit baseline.
Thus, the final uncertainties on the corrected transmission spectra are limited by the transit duration and not by the out-of-transit baseline.
By contrast, for the cross-retrievals, $\sigma(\epsilon)/\epsilon$ is comparable to $\sigma(D)/D$ for all baselines because the model fidelity limits the precision of the photospheric inferences, not the data uncertainty.
From this we conclude that when aiming to correct for the TLS effect using a model-based approach, an out-of-transit baseline that is at least double the transit duration is sufficient to shift the limiting factor in the final precision from the data to the models.
However, empirical approaches to TLS corrections \citep[e.g.,][]{TRAPPIST-1JWSTCommunityInitiative2023, Berardo2024} and other considerations, such as stellar variability, will warrant longer out-of-transit baselines in some cases.
}

\begin{figure*}
    \centering
    \includegraphics[width=\textwidth]{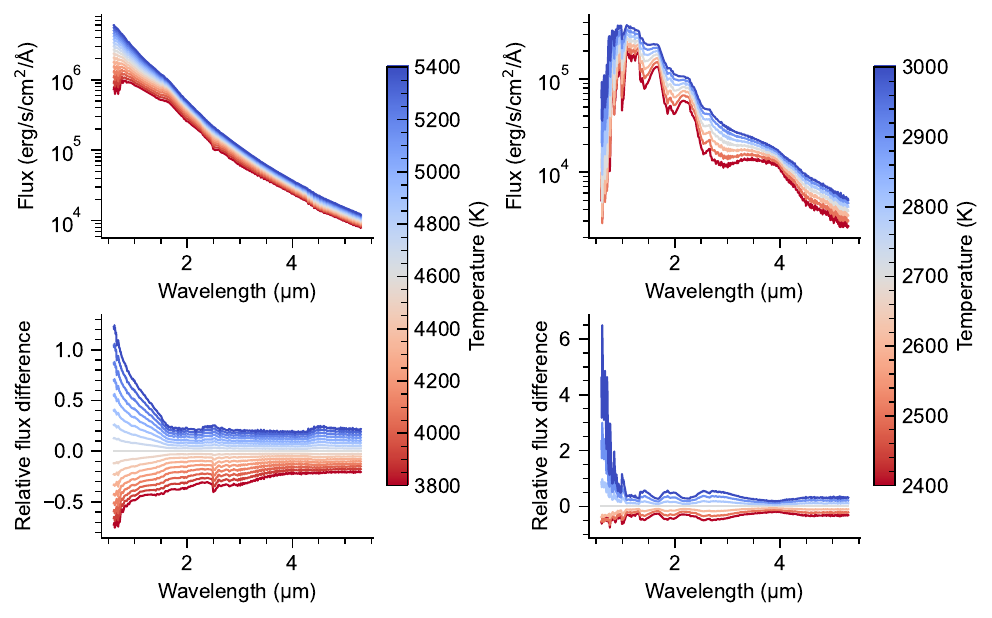}
    \caption{
    Flux changes due to temperature variations for the range of stellar models we consider.
    The left column shows PHOENIX models relevant to our K0 case, and the right column shows PHOENIX models relevant to our M6 case.
    The top panels show the spectra in absolute flux units, while the bottom panels show each set normalized to the middle-temperature model in each set.
    The wavelengths and resolution of the spectra are relevant to NIRSpec/PRISM.
    Larger flux differences are evident for the set of models relevant to the M6 case, which lead to more successful inferences from the retrievals.
    \label{fig:temp_variations}
    }
\end{figure*}

\section{Conclusions \& Future Work} \label{sec:discussion}

We investigated the use of out-of-transit stellar spectra to enhance \JWST{}'s scientific return while reducing biases in exoplanet transmission spectra, with a focus on the impact of stellar model fidelity. 
Our analysis produced two primary findings.
\begin{enumerate}
    \item The fidelity of stellar models is crucially important for identifying the right complexity of a photosphere and deriving appropriate corrections for transmission spectra.
    The differences between existing model grids dominate the total noise budget. 
    This translates into needing to inflate photon-noise error bars by \R{50\% or more} (\autoref{fig:relative_uncertainties}), which prevents efforts from harnessing the full potential of \JWST{} for transits of stars with heterogeneous photospheres. 
    \R{Moreover, we} note that even when accounting for this inflation, significant biases on the derived properties of the stellar photosphere are possible, leading to improper corrections \R{(\autoref{fig:bias_reduction}, bottom right panel)}. 
    This finding is similar to earlier findings of \citet{deWit2012} and \citetalias{Niraula2022}, which have shown that an apparently good fit can hide a compensation for a model's lack of fidelity via biases in the model parameters.
    
    \item If the model fidelity is on par with the precision of the spectra, it is possible to reliably infer the correct model parameters (including the true number of components). 
    This means that with sufficient model fidelity\R{---meaning that the relative error on the stellar model is smaller than the relative uncertainty on the transit depth (e.g., 2.5\,ppt for a 2\% transit depth measured at 50\,ppm)---}one can expect to correct for stellar contamination to the maximum extent possible, given the information content of the data, and there is no model-driven bottleneck. 
    In this context, we show that \R{as long as the out-of-transit baseline is at least twice the transit duration,} the uncertainty associated with the correction of the stellar contamination is marginal compared to the photon noise on the transmission spectrum, thereby allowing photon-limited science (i.e., harnessing the full potential of \JWST{}).
\end{enumerate}

These findings should motivate both further theoretical developments in modeling the spectra of stars and their heterogeneities and new \R{observational} strategies to derive empirical constraints on stellar photospheric heterogeneity from highly precise \JWST{} spectra.
In fact, we suggest that observational strategies could be developed to acquire these empirical constraints with the observatories with which the planetary atmospheres will be explored to ensure a ``fidelity'' on par with the data driving the atmospheric characterization.
We provide in the following a few additional considerations for future works.

\subsection{Spectral Model Grids and Interpolation Schemes}

\R{The typical standard deviation of our component temperature estimates is 4.5\,K (\autoref{sec:parameter_inference})}, while the spacing of the temperature grids is 100\,K for all three model grids used. 
This means that the sampling of the model grid is insufficient for the high-S/N data at hand.
\R{In other words, while precise \JWST{} may be able to constrain the temperatures of spectral components at the ${\sim}5$\,K level, this effort is limited by the relative sparse sampling of 100\,K in available grids (roughly 20 times greater than possible constraints.)}
\R{To} support the reliable correction of stellar contamination in \JWST{} exoplanet transmission spectra, we suggest that it would be useful to generate model grids with spacings in \R{temperature (and other dimensions, where possible)} that are two orders of magnitude smaller than those currently available \R{(e.g., 1\,K) in order to bring the model spacing in line with constraints that are now possible with \JWST{} spectra.}

In addition, \R{for simplicity} we use a linear interpolation scheme \R{here}, which could also contribute to reducing the fidelity of the models over the coarse grid available \citep[see, e.g.,][]{Czekala2015}.
We also recommend that future work explore the impact in this context of linear interpolation versus more \R{precise (but complex)} approaches, such as bicubic interpolation or spectral emulation via principal component analysis \citep[e.g.,][]{Czekala2015} \R{or machine learning \citep{Badenas-Agusti2024}}.

\subsection{Heterogeneities Are Not Your Average Photosphere}

Out-of-transit spectra are currently fitted using a combination of stellar spectra weighted by different filling factors. 
This approach thus assumes that spot and facula spectra can be approximated by stellar spectra derived from 1D radiative--convective models. 
Although this assumption may be passable for spots \citep{Rackham2023}, it has been shown to be a poor assumption for facula spectra, which contain magnetically induced features that are not captured well by the simplified 1D models \citep{Witzke2022}. 
The increased differentiation of a component's features, while more problematic because the components are more challenging to approximate with current models, will also make their contribution easier to disentangle in observations. 
While a new generation of calculations for the spectra of heterogeneities are underway \citep[e.g., with MPS-ATLAS;][]{Witzke2021}, the prospect of supporting the benchmarking of said models with empirical constraints within reach with \JWST{} is tantalizing.

\subsection{When Worse Can Also Mean Better}

\begin{figure}
    \centering
    \includegraphics[width=\columnwidth]{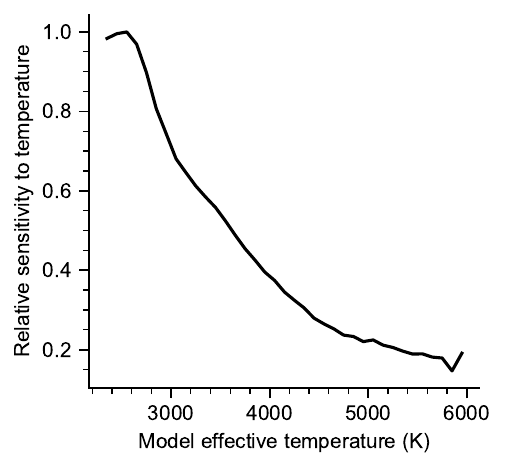}
    \caption{
    The sensitivity of PHOENIX stellar models, sampled at the wavelength range and resolution NIRSpec/PRISM, to changes in stellar temperature.
    The relative change in stellar flux is shown as a function of the model effective temperature, normalized to the value for a model of 2500\,K.
    The sensitivity to temperature decreases with increasing temperature, with models around 6000\,K displaying roughly 20\% of the sensitivity of the 2500\,K model.
    \label{fig:temp_sensitivity}
    }
\end{figure}

As with facula spectra, for which the challenge of fitting them is also an unexpected benefit, the challenge of constraining photospheric heterogeneity in cooler stars may be lessened by that same heterogeneity.
In \autoref{sec:corrections} we found that\R{, before applying an achromatic offset, the corrections} derived from the direct \R{retrievals} of the K0 \texttt{case 3l} \R{and, to a lesser extent, M6 \texttt{case 3h} spectra} actually increased the bias in the transmission \R{spectra (\autoref{fig:bias_reduction})}, due to \R{multimodal posterior distributions} derived from the out-of-transit stellar \R{spectra}.
Ultimately, these incorrect inferences, deriving from high-S/N spectra simulated and fitted with the same spectral grid, highlight the challenge of deriving constraints in this temperature regime.
\autoref{fig:temp_variations} shows the sensitivity of stellar spectra to temperature variations over ranges covering the components of our K0 and M6 cases.
It highlights that the sensitivity of the spectra relevant to a K0 and its photospheric components is smaller than for the M6 case (\autoref{fig:temp_sensitivity}), which relates to the expectation of a lower level of stellar contamination. 
However, the \R{strengths} of temperature-sensitive features in the spectra actually support the detection and characterization of these heterogeneities, and thus the correction of stellar contamination. 
Thus, we note that the lack of significant differentiation in the K0 models can also lead to biased inferences when a particular noise realization permits another nearby, nearly blackbody model to fit the data, though working at higher resolving power with other \JWST{} observational modes (e.g., NIRSpec/G395H, NIRISS SOSS) \R{may} help here \R{and is the topic of a future study}.

\R{\subsection{Mind the Gaps: Carefully Navigating Multimodality}}

\R{Due to an increase in information content for \JWST{} spectra, the posterior probability distributions of derived photospheric parameters are tighter than found with even \HST{} spectra \citep[e.g.,][]{Garcia2022} and may exhibit multiple modes (i.e., solutions) that would otherwise be indistinguishable. 
In other words, the increase in instrumental precision leads to a more granular distribution of probability (or $\chi^2$) over the parameter space. 
Additionally, as we are increasingly able to probe contributions from multiple spectral components, the resulting increase in the dimensionality of the parameter space may also increase the likelihood of modes emerging.
As the field of exoplanetary science moves into an era of high information content data, we expect to increasingly see multimodal parameter estimates for stellar photospheric properties and concomitant stellar contamination signals.
}

\R{Multimodal distributions are challenging to navigate with classical analysis techniques like Markov Chain Monte Carlo algorithms, which tend to home in on one local minimum and may not identify other possible solutions. 
Nested sampling approaches, as used here, can more effectively identify multimodal posteriors \citep{Buchner2014, Buchner2017}.
However, as reported in \autoref{sec:parameter_inference}, significant biases in the inferences are still possible with this approach. 
Even if the minimum associated with truth is probed, the mean parameter estimate may be significantly biased from the truth by the inclusion of many other modes in the posterior sample.
In turn, this results in stellar contamination corrections that increase the uncertainty on the transmission spectrum but do not reduce its bias.
In this context, we suggest that special care be taken in future works to further develop and refine approaches for handling multimodal distributions in the context of parameter estimates for photospheric properties of exoplanet hosts.
}

\subsection{Empirical Heterogeneity Constraints}

We have worked under the assumption that different stellar model grids are equally good.
Another possibility is that advances, including, among other things, updated opacities for a wide range of sources \citep[e.g.,][]{Tennyson2016, Tennyson2020, Gordon2017, Gordon2022}, have allowed more recent grids to more closely resemble reality.
In this case, the differences between model grids then reflect the growth in our understanding rather than a remaining understanding gap to cross.
To assess this possibility, we recommend that these techniques be applied to real \JWST{} data, starting with an inactive star and advancing to more active stars to understand the limits of model-based inferences with real data---keeping in mind that a ``good fit'' does not automatically imply ``model fidelity.''
We also recommend the \R{further} exploration of empirical approaches for deriving the unique spectral components of a photosphere and their filling factors \R{\citep[e.g.,][]{Berardo2024}}, enabling corrections that are independent of spectral models.

\vspace{1.5ex}
\section*{Acknowledgements}
\R{We thank the reviewer for the constructive feedback that has improved this study.}
We thank Prajwal Niraula for providing the transmission spectra from \citetalias{Niraula2022}.
We also thank Aisha Iyer for providing the SPHINX spectral grid \citep{Iyer2023} and Sasha Shapiro and Nadiia Kostogryz for pointing us to the MPS-ATLAS model library \citep{Witzke2021}.
B.V.R. thanks the Heising-Simons Foundation for support.
This material is based upon work supported by the National Aeronautics and Space Administration under Agreement No.\ 80NSSC21K0593 for the program ``Alien Earths.''
The results reported herein benefited from collaborations and/or information exchange within NASA's Nexus for Exoplanet System Science (NExSS) research coordination network sponsored by NASA's Science Mission Directorate.
The authors acknowledge the MIT SuperCloud and Lincoln Laboratory Supercomputing Center for providing (HPC, database, consultation) resources that have contributed to the research results reported within this paper/report.
\R{
This research has made use of NASA's Astrophysics Data System Bibliographic Services.
This research has made use of the SIMBAD database, CDS, Strasbourg Astronomical Observatory, France \citep{Simbad}.
This research has made use of the NASA Exoplanet Archive, which is operated by the California Institute of Technology, under contract with the National Aeronautics and Space Administration under the Exoplanet Exploration Program.
}

%

\vspace{5mm}
\facilities{Exoplanet Archive.}


\software{
\texttt{Astropy} \citep{AstropyCollaboration2013, AstropyCollaboration2018, AstropyCollaboration2022}, \texttt{Matplotlib} \citep{Hunter2007}, 
\texttt{NumPy} \citep{NumPy}, 
\texttt{Pandas} \citep{pandas},
\texttt{SciPy} \citep{SciPy}, 
\texttt{Speclib} \citep{speclib-0.0-beta.0}
}



\appendix

\section{Goodness-of-fit results}
\label{appendix:gof}

\R{
Tables~\ref{tab:gof_results_direct} and \ref{tab:gof_results_cross} provide the full goodness-of-fit details for the direct- and cross-retrieval fits, respectively.
For simplicity, we focus on the results from the first of the five noise instances tested.
For each retrieval, these tables provide the degrees of freedom (dof), the $\chi^2$ value, the $\chi^2$ value after accounting for the error inflation determined via the fit ($\chi^2_\mathrm{infl}$), the reduced $\chi^2$ values before and after accounting for the error inflation ($\chi^2_r$ and $\chi^2_{r, \mathrm{infl}}$, respectively), the marginal Bayesian evidence ($\Delta \ln \mathcal{Z}$), the marginal Bayesian information criterion ($\Delta$BIC), the marginal AIC ($\Delta$AIC), the Akaike weight ($w$), and its accompanying evidence ratio (ER).
$\Delta \ln \mathcal{Z}$ is defined based on the Bayesian-evidence-based criteria presented in \autoref{sec:model_selection}, and $\Delta$BIC and $\Delta$AIC are defined with respect to the minimum BIC and AIC, respectively, in a set of models. 
}

\begin{deluxetable}{cccCCCCCCCCCCC}
\tablecaption{Goodness-of-fit Results for Direct Retrievals with the First Noise Instance \label{tab:gof_results_direct}}
\tablehead{\colhead{Grid} & \colhead{Star} & \colhead{Case} & \colhead{Model} & \colhead{dof} & \colhead{$\chi^2$} & \colhead{$\chi^2_\mathrm{infl}$} & \colhead{$\chi^2_r$} & \colhead{$\chi^2_{r, \mathrm{infl}}$} & \colhead{$\Delta \ln \mathcal{Z}$} & \colhead{$\Delta$BIC} & \colhead{$\Delta$AIC} & \colhead{$w$} & \colhead{ER}}
\startdata
PHOENIX & K0 & $\texttt{1}$ & 1 & 400 & 382 & 382 & 0.955 & 0.954 & 0 & 0 & 0 & 0.946 & 1.00 \\
PHOENIX & K0 & $\texttt{1}$ & 2 & 397 & 382 & 380 & 0.962 & 0.957 & -7.86 & 18.0 & 6.00 & 0.0471 & 20.1 \\
PHOENIX & K0 & $\texttt{1}$ & 3 & 395 & 382 & 380 & 0.967 & 0.961 & -12.7 & 30.0 & 10.0 & 6.33\times10^{-3} & 149 \\
PHOENIX & K0 & $\texttt{1}$ & 4 & 393 & 382 & 380 & 0.972 & 0.966 & -18.4 & 42.0 & 14.0 & 8.62\times10^{-4} & ${>}10^3$ \\
PHOENIX & K0 & $\texttt{2l}$ & 1 & 400 & 1.17\times10^5 & 399 & 293 & 0.998 & -902 & ${>}10^3$ & ${>}10^3$ & 0 & ${>}10^3$ \\
PHOENIX & K0 & $\texttt{2l}$ & 2 & 397 & 366 & 366 & 0.922 & 0.922 & 0 & 0 & 0 & 0.882 & 1.00 \\
PHOENIX & K0 & $\texttt{2l}$ & 3 & 395 & 366 & 359 & 0.927 & 0.908 & -7.93 & 12.3 & 4.31 & 0.102 & 8.64 \\
PHOENIX & K0 & $\texttt{2l}$ & 4 & 393 & 366 & 360 & 0.932 & 0.916 & -11.4 & 24.0 & 8.04 & 0.0159 & 55.6 \\
PHOENIX & K0 & $\texttt{2h}$ & 1 & 400 & 3.13\times10^6 & 399 & 7.83\times10^3 & 0.998 & -1.52\times10^3 & ${>}10^3$ & ${>}10^3$ & 0 & ${>}10^3$ \\
PHOENIX & K0 & $\texttt{2h}$ & 2 & 397 & 417 & 413 & 1.05 & 1.04 & 0 & 0 & 0 & 0.867 & 1.00 \\
PHOENIX & K0 & $\texttt{2h}$ & 3 & 395 & 417 & 414 & 1.06 & 1.05 & -2.53 & 12.0 & 3.99 & 0.118 & 7.37 \\
PHOENIX & K0 & $\texttt{2h}$ & 4 & 393 & 417 & 410 & 1.06 & 1.04 & -6.49 & 24.0 & 8.04 & 0.0156 & 55.6 \\
PHOENIX & K0 & $\texttt{3l}$ & 1 & 400 & 1.20\times10^5 & 401 & 300 & 1.00 & -862 & {>}10^3 & {>}10^3 & 0 & ${>}10^3$ \\
PHOENIX & K0 & $\texttt{3l}$ & 2 & 397 & 469 & 389 & 1.18 & 0.98 & -11.7 & 22.3 & 30.3 & {<}10^{-6} & ${>}10^3$ \\
PHOENIX & K0 & $\texttt{3l}$ & 3 & 395 & 426 & 388 & 1.08 & 0.983 & 0 & 0 & 0 & 0.879 & 1.00 \\
PHOENIX & K0 & $\texttt{3l}$ & 4 & 393 & 426 & 373 & 1.08 & 0.95 & -1.73 & 12.0 & 3.96 & 0.121 & 7.26 \\
PHOENIX & K0 & $\texttt{3h}$ & 1 & 400 & 3.15\times10^6 & 400 & 7.87\times10^3 & 0.999 & -1.51\times10^3 & ${>}10^3$ & ${>}10^3$ & 0 & ${>}10^3$ \\
PHOENIX & K0 & $\texttt{3h}$ & 2 & 397 & 723 & 477 & 1.82 & 1.20 & -106 & 217 & 225 & {<}10^{-6} & ${>}10^3$ \\
PHOENIX & K0 & $\texttt{3h}$ & 3 & 395 & 423 & 420 & 1.07 & 1.06 & 0 & 0 & 0 & 1 & 1.00 \\
PHOENIX & K0 & $\texttt{3h}$ & 4 & 393 & 451 & 446 & 1.15 & 1.14 & -16.7 & 38.2 & 30.2 & {<}10^{-6} & ${>}10^3$ \\
PHOENIX & M6 & $\texttt{1}$ & 1 & 400 & 466 & 457 & 1.17 & 1.14 & 0 & 0 & 0 & 0.936 & 1.00 \\
PHOENIX & M6 & $\texttt{1}$ & 2 & 397 & 466 & 436 & 1.17 & 1.10 & -7.30 & 17.7 & 5.66 & 0.0552 & 17.0 \\
PHOENIX & M6 & $\texttt{1}$ & 3 & 395 & 466 & 427 & 1.18 & 1.08 & -12.3 & 29.6 & 9.65 & 7.53\times10^{-3} & 124 \\
PHOENIX & M6 & $\texttt{1}$ & 4 & 393 & 466 & 433 & 1.19 & 1.1 & -18.2 & 41.7 & 13.7 & 9.72\times10^{-4} & 963 \\
PHOENIX & M6 & $\texttt{2l}$ & 1 & 400 & 2.63\times10^4 & 424 & 65.8 & 1.06 & -713 & ${>}10^3$ & ${>}10^3$ & {<}10^{-6} & ${>}10^3$ \\
PHOENIX & M6 & $\texttt{2l}$ & 2 & 397 & 389 & 389 & 0.980 & 0.979 & 0 & 0 & 0 & 0.551 & 1.00 \\
PHOENIX & M6 & $\texttt{2l}$ & 3 & 395 & 387 & 383 & 0.979 & 0.97 & -6.02 & 8.49 & 0.492 & 0.431 & 1.28 \\
PHOENIX & M6 & $\texttt{2l}$ & 4 & 393 & 389 & 384 & 0.989 & 0.977 & -10.3 & 22.8 & 6.79 & 0.0185 & 29.7 \\
PHOENIX & M6 & $\texttt{2h}$ & 1 & 400 & 7.98\times10^5 & 403 & 2.00\times10^3 & 1.01 & -1.37\times10^3 & ${>}10^3$ & ${>}10^3$ & 0 & ${>}10^3$ \\
PHOENIX & M6 & $\texttt{2h}$ & 2 & 397 & 372 & 371 & 0.936 & 0.934 & 0 & 0 & 0 & 0.867 & 1.00 \\
PHOENIX & M6 & $\texttt{2h}$ & 3 & 395 & 372 & 366 & 0.941 & 0.926 & -4.57 & 12.0 & 4.01 & 0.117 & 7.44 \\
PHOENIX & M6 & $\texttt{2h}$ & 4 & 393 & 372 & 366 & 0.946 & 0.932 & -9.69 & 24.0 & 7.98 & 0.016 & 54.1 \\
PHOENIX & M6 & $\texttt{3l}$ & 1 & 400 & 3.60\times10^4 & 427 & 90.1 & 1.07 & -773 & ${>}10^3$ & ${>}10^3$ & 0 & ${>}10^3$ \\
PHOENIX & M6 & $\texttt{3l}$ & 2 & 397 & 405 & 404 & 1.02 & 1.02 & 0 & 0 & 6.76 & 0.0303 & 29.4 \\
PHOENIX & M6 & $\texttt{3l}$ & 3 & 395 & 395 & 392 & 1.00 & 0.992 & -6.22 & 1.24 & 0 & 0.891 & 1.00 \\
PHOENIX & M6 & $\texttt{3l}$ & 4 & 393 & 395 & 394 & 1.00 & 1.00 & -3.69 & 14.1 & 4.85 & 0.0787 & 11.3 \\
PHOENIX & M6 & $\texttt{3h}$ & 1 & 400 & 9.04\times10^5 & 404 & 2.26\times10^3 & 1.01 & -1.38\times10^3 & ${>}10^3$ & ${>}10^3$ & 0 & ${>}10^3$ \\
PHOENIX & M6 & $\texttt{3h}$ & 2 & 397 & 534 & 484 & 1.35 & 1.22 & -55.8 & 111 & 119 & {<}10^{-6} & ${>}10^3$ \\
PHOENIX & M6 & $\texttt{3h}$ & 3 & 395 & 406 & 405 & 1.03 & 1.02 & 0 & 0 & 0 & 0.913 & 1.00 \\
PHOENIX & M6 & $\texttt{3h}$ & 4 & 393 & 407 & 391 & 1.03 & 0.994 & -9.63 & 12.7 & 4.71 & 0.0868 & 10.5
\enddata
\end{deluxetable}

\begin{deluxetable}{cccCCCCCCCCCCC}
\tablecaption{Goodness-of-fit Results for Cross-retrievals with the First Noise Instance \label{tab:gof_results_cross}}
\tablehead{\colhead{Grid} & \colhead{Star} & \colhead{Case} & \colhead{Model} & \colhead{dof} & \colhead{$\chi^2$} & \colhead{$\chi^2_\mathrm{infl}$} & \colhead{$\chi^2_r$} & \colhead{$\chi^2_{r, \mathrm{infl}}$} & \colhead{$\Delta \ln \mathcal{Z}$} & \colhead{$\Delta$BIC} & \colhead{$\Delta$AIC} & \colhead{$w$} & \colhead{ER}}
\startdata
MPS-ATLAS & K0 & $\texttt{1}$ & 1 & 400 & 5.92\times10^6 & 400 & 1.48\times10^4 & 1.00 & -60.0 & 118 & 130 & {<}10^{-6} & {>}10^{3} \\
MPS-ATLAS & K0 & $\texttt{1}$ & 2 & 397 & 7.24\times10^6 & 398 & 1.82\times10^4 & 1.00 & 0 & 0 & 0 & 0.827 & 1 \\
MPS-ATLAS & K0 & $\texttt{1}$ & 3 & 395 & 7.18\times10^6 & 398 & 1.82\times10^4 & 1.01 & -1.34 & 11.4 & 3.38 & 0.152 & 5.43 \\
MPS-ATLAS & K0 & $\texttt{1}$ & 4 & 393 & 7.12\times10^6 & 397 & 1.81\times10^4 & 1.01 & -4.16 & 23.3 & 7.33 & 0.0211 & 39.1 \\
MPS-ATLAS & K0 & $\texttt{2l}$ & 1 & 400 & 6.83\times10^6 & 400 & 1.71\times10^4 & 1.00 & -75.4 & 149 & 161 & {<}10^{-6} & {>}10^{3} \\
MPS-ATLAS & K0 & $\texttt{2l}$ & 2 & 397 & 7.17\times10^6 & 398 & 1.81\times10^4 & 1.00 & 0 & 0 & 0 & 0.815 & 1 \\
MPS-ATLAS & K0 & $\texttt{2l}$ & 3 & 395 & 7.15\times10^6 & 398 & 1.81\times10^4 & 1.01 & -0.771 & 11.2 & 3.23 & 0.162 & 5.03 \\
MPS-ATLAS & K0 & $\texttt{2l}$ & 4 & 393 & 7.12\times10^6 & 398 & 1.81\times10^4 & 1.01 & -3.09 & 23.2 & 7.16 & 0.0227 & 35.9 \\
MPS-ATLAS & K0 & $\texttt{2h}$ & 1 & 400 & 2.33\times10^7 & 402 & 5.81\times10^4 & 1.00 & -168 & 329 & 341 & {<}10^{-6} & {>}10^{3} \\
MPS-ATLAS & K0 & $\texttt{2h}$ & 2 & 397 & 7.15\times10^6 & 397 & 1.8\times10^4 & 0.999 & 0 & 0 & 0 & 0.752 & 1 \\
MPS-ATLAS & K0 & $\texttt{2h}$ & 3 & 395 & 7.11\times10^6 & 398 & 1.8\times10^4 & 1.01 & -1.06 & 10.5 & 2.50 & 0.216 & 3.48 \\
MPS-ATLAS & K0 & $\texttt{2h}$ & 4 & 393 & 7.12\times10^6 & 399 & 1.81\times10^4 & 1.02 & -3.51 & 22.3 & 6.32 & 0.0319 & 23.5 \\
MPS-ATLAS & K0 & $\texttt{3l}$ & 1 & 400 & 6.64\times10^6 & 399 & 1.66\times10^4 & 0.998 & -74.5 & 146 & 158 & {<}10^{-6} & {>}10^{3} \\
MPS-ATLAS & K0 & $\texttt{3l}$ & 2 & 397 & 7.15\times10^6 & 398 & 1.8\times10^4 & 1.00 & 0 & 0 & 0 & 0.802 & 1 \\
MPS-ATLAS & K0 & $\texttt{3l}$ & 3 & 395 & 7.07\times10^6 & 398 & 1.79\times10^4 & 1.01 & -1.15 & 11.1 & 3.05 & 0.174 & 4.60 \\
MPS-ATLAS & K0 & $\texttt{3l}$ & 4 & 393 & 7.02\times10^6 & 398 & 1.79\times10^4 & 1.01 & -4.08 & 23.0 & 7.05 & 0.0236 & 34.0 \\
MPS-ATLAS & K0 & $\texttt{3h}$ & 1 & 400 & 1.25\times10^7 & 399 & 3.13\times10^4 & 0.997 & -138 & 273 & 285 & {<}10^{-6} & {>}10^{3} \\
MPS-ATLAS & K0 & $\texttt{3h}$ & 2 & 397 & 6.98\times10^6 & 399 & 1.76\times10^4 & 1.00 & 0 & 0 & 0 & 0.646 & 1 \\
MPS-ATLAS & K0 & $\texttt{3h}$ & 3 & 395 & 6.88\times10^6 & 398 & 1.74\times10^4 & 1.01 & -1.43 & 9.45 & 1.45 & 0.312 & 2.07 \\
MPS-ATLAS & K0 & $\texttt{3h}$ & 4 & 393 & 6.86\times10^6 & 400 & 1.74\times10^4 & 1.02 & -3.39 & 21.4 & 5.44 & 0.0426 & 15.2 \\
SPHINX & M6 & $\texttt{1}$ & 1 & 400 & 1.12\times10^8 & 400 & 2.81\times10^5 & 1.00 & -18.6 & 40.1 & 60.1 & {<}10^{-6} & {>}10^{3} \\
SPHINX & M6 & $\texttt{1}$ & 2 & 397 & 1.15\times10^8 & 399 & 2.89\times10^5 & 1.01 & 0 & 0.819 & 8.82 & 0.0102 & 82.1 \\
SPHINX & M6 & $\texttt{1}$ & 3 & 395 & 1.28\times10^8 & 399 & 3.23\times10^5 & 1.01 & 3.78 & 0 & 0 & 0.837 & 1 \\
SPHINX & M6 & $\texttt{1}$ & 4 & 393 & 1.28\times10^8 & 400 & 3.27\times10^5 & 1.02 & 2.45 & 11.4 & 3.41 & 0.152 & 5.49 \\
SPHINX & M6 & $\texttt{2l}$ & 1 & 400 & 1.13\times10^8 & 399 & 2.82\times10^5 & 0.997 & -19.0 & 43.0 & 62.9 & {<}10^{-6} & {>}10^{3} \\
SPHINX & M6 & $\texttt{2l}$ & 2 & 397 & 1.14\times10^8 & 400 & 2.87\times10^5 & 1.01 & 0 & 2.65 & 10.6 & 4.16\times10^{-3} & 205 \\
SPHINX & M6 & $\texttt{2l}$ & 3 & 395 & 1.27\times10^8 & 399 & 3.22\times10^5 & 1.01 & 4.54 & 0 & 0 & 0.853 & 1 \\
SPHINX & M6 & $\texttt{2l}$ & 4 & 393 & 1.28\times10^8 & 399 & 3.25\times10^5 & 1.02 & 2.76 & 11.6 & 3.57 & 0.143 & 5.97 \\
SPHINX & M6 & $\texttt{2h}$ & 1 & 400 & 1.14\times10^8 & 399 & 2.85\times10^5 & 0.998 & -27.7 & 53.3 & 73.3 & {<}10^{-6} & {>}10^{3} \\
SPHINX & M6 & $\texttt{2h}$ & 2 & 397 & 1.12\times10^8 & 400 & 2.83\times10^5 & 1.01 & -8.02 & 10.3 & 18.3 & 8.68\times10^{-5} & {>}10^{3} \\
SPHINX & M6 & $\texttt{2h}$ & 3 & 395 & 1.26\times10^8 & 399 & 3.18\times10^5 & 1.01 & 0 & 0 & 0 & 0.81 & 1 \\
SPHINX & M6 & $\texttt{2h}$ & 4 & 393 & 1.27\times10^8 & 399 & 3.22\times10^5 & 1.02 & -1.51 & 10.9 & 2.90 & 0.19 & 4.26 \\
SPHINX & M6 & $\texttt{3l}$ & 1 & 400 & 1.05\times10^8 & 399 & 2.63\times10^5 & 0.997 & -22.4 & 48.1 & 63.5 & {<}10^{-6} & {>}10^{3} \\
SPHINX & M6 & $\texttt{3l}$ & 2 & 397 & 1.09\times10^8 & 400 & 2.75\times10^5 & 1.01 & 0 & 0 & 3.40 & 0.139 & 5.48 \\
SPHINX & M6 & $\texttt{3l}$ & 3 & 395 & 1.19\times10^8 & 399 & 3.02\times10^5 & 1.01 & 1.94 & 4.60 & 0 & 0.761 & 1 \\
SPHINX & M6 & $\texttt{3l}$ & 4 & 393 & 1.21\times10^8 & 399 & 3.07\times10^5 & 1.02 & -0.183 & 16.7 & 4.06 & 0.1 & 7.61 \\
SPHINX & M6 & $\texttt{3h}$ & 1 & 400 & 9.33\times10^7 & 401 & 2.33\times10^5 & 1.00 & -32.0 & 66.2 & 78.2 & {<}10^{-6} & {>}10^{3} \\
SPHINX & M6 & $\texttt{3h}$ & 2 & 397 & 9.95\times10^7 & 400 & 2.51\times10^5 & 1.01 & 0 & 0 & 0 & 0.756 & 1 \\
SPHINX & M6 & $\texttt{3h}$ & 3 & 395 & 1.04\times10^8 & 399 & 2.64\times10^5 & 1.01 & -0.429 & 10.5 & 2.51 & 0.216 & 3.51 \\
SPHINX & M6 & $\texttt{3h}$ & 4 & 393 & 1.06\times10^8 & 398 & 2.69\times10^5 & 1.01 & -3.21 & 22.6 & 6.59 & 0.0280 & 27.0
\enddata
\end{deluxetable}

\bibliography{bibliography, local}{}

\begin{thebibliography}{}
\expandafter\ifx\csname natexlab\endcsname\relax\def\natexlab#1{#1}\fi
\providecommand{\url}[1]{\href{#1}{#1}}
\providecommand{\dodoi}[1]{doi:~\href{http://doi.org/#1}{\nolinkurl{#1}}}
\providecommand{\doeprint}[1]{\href{http://ascl.net/#1}{\nolinkurl{http://ascl.net/#1}}}
\providecommand{\doarXiv}[1]{\href{https://arxiv.org/abs/#1}{\nolinkurl{https://arxiv.org/abs/#1}}}

\bibitem[{{Afram} \& {Berdyugina}(2015)}]{Afram2015}
{Afram}, N., \& {Berdyugina}, S.~V. 2015, \aap, 576, A34,
  \dodoi{10.1051/0004-6361/201425314}

\bibitem[{{Ahrer} {et~al.}(2023){Ahrer}, {Stevenson}, {Mansfield}, {Moran},
  {Brande}, {Morello}, {Murray}, {Nikolov}, {Petit dit de la Roche},
  {Schlawin}, {Wheatley}, {Zieba}, {Batalha}, {Damiano}, {Goyal}, {Lendl},
  {Lothringer}, {Mukherjee}, {Ohno}, {Batalha}, {Battley}, {Bean}, {Beatty},
  {Benneke}, {Berta-Thompson}, {Carter}, {Cubillos}, {Daylan}, {Espinoza},
  {Gao}, {Gibson}, {Gill}, {Harrington}, {Hu}, {Kreidberg}, {Lewis}, {Line},
  {L{\'o}pez-Morales}, {Parmentier}, {Powell}, {Sing}, {Tsai}, {Wakeford},
  {Welbanks}, {Alam}, {Alderson}, {Allen}, {Anderson}, {Barstow}, {Bayliss},
  {Bell}, {Blecic}, {Bryant}, {Burleigh}, {Carone}, {Casewell}, {Changeat},
  {Chubb}, {Crossfield}, {Crouzet}, {Decin}, {D{\'e}sert}, {Feinstein},
  {Flagg}, {Fortney}, {Gizis}, {Heng}, {Iro}, {Kempton}, {Kendrew}, {Kirk},
  {Knutson}, {Komacek}, {Lagage}, {Leconte}, {Lustig-Yaeger}, {MacDonald},
  {Mancini}, {May}, {Mayne}, {Miguel}, {Mikal-Evans}, {Molaverdikhani},
  {Palle}, {Piaulet}, {Rackham}, {Redfield}, {Rogers}, {Roy}, {Rustamkulov},
  {Shkolnik}, {Sotzen}, {Taylor}, {Tremblin}, {Tucker}, {Turner}, {de
  Val-Borro}, {Venot}, \& {Zhang}}]{Ahrer2023}
{Ahrer}, E.-M., {Stevenson}, K.~B., {Mansfield}, M., {et~al.} 2023, \nat, 614,
  653, \dodoi{10.1038/s41586-022-05590-4}

\bibitem[{{Akaike}(1974)}]{Akaike1974}
{Akaike}, H. 1974, IEEE Transactions on Automatic Control, 19, 716

\bibitem[{{Alderson} {et~al.}(2023){Alderson}, {Wakeford}, {Alam}, {Batalha},
  {Lothringer}, {Adams Redai}, {Barat}, {Brande}, {Damiano}, {Daylan},
  {Espinoza}, {Flagg}, {Goyal}, {Grant}, {Hu}, {Inglis}, {Lee}, {Mikal-Evans},
  {Ramos-Rosado}, {Roy}, {Wallack}, {Batalha}, {Bean}, {Benneke},
  {Berta-Thompson}, {Carter}, {Changeat}, {Col{\'o}n}, {Crossfield},
  {D{\'e}sert}, {Foreman-Mackey}, {Gibson}, {Kreidberg}, {Line},
  {L{\'o}pez-Morales}, {Molaverdikhani}, {Moran}, {Morello}, {Moses},
  {Mukherjee}, {Schlawin}, {Sing}, {Stevenson}, {Taylor}, {Aggarwal}, {Ahrer},
  {Allen}, {Barstow}, {Bell}, {Blecic}, {Casewell}, {Chubb}, {Crouzet},
  {Cubillos}, {Decin}, {Feinstein}, {Fortney}, {Harrington}, {Heng}, {Iro},
  {Kempton}, {Kirk}, {Knutson}, {Krick}, {Leconte}, {Lendl}, {MacDonald},
  {Mancini}, {Mansfield}, {May}, {Mayne}, {Miguel}, {Nikolov}, {Ohno}, {Palle},
  {Parmentier}, {Petit dit de la Roche}, {Piaulet}, {Powell}, {Rackham},
  {Redfield}, {Rogers}, {Rustamkulov}, {Tan}, {Tremblin}, {Tsai}, {Turner}, {de
  Val-Borro}, {Venot}, {Welbanks}, {Wheatley}, \& {Zhang}}]{Alderson2023}
{Alderson}, L., {Wakeford}, H.~R., {Alam}, M.~K., {et~al.} 2023, \nat, 614,
  664, \dodoi{10.1038/s41586-022-05591-3}

\bibitem[{{Astropy Collaboration} {et~al.}(2013){Astropy Collaboration},
  {Robitaille}, {Tollerud}, {Greenfield}, {Droettboom}, {Bray}, {Aldcroft},
  {Davis}, {Ginsburg}, {Price-Whelan}, {Kerzendorf}, {Conley}, {Crighton},
  {Barbary}, {Muna}, {Ferguson}, {Grollier}, {Parikh}, {Nair}, {Unther},
  {Deil}, {Woillez}, {Conseil}, {Kramer}, {Turner}, {Singer}, {Fox}, {Weaver},
  {Zabalza}, {Edwards}, {Azalee Bostroem}, {Burke}, {Casey}, {Crawford},
  {Dencheva}, {Ely}, {Jenness}, {Labrie}, {Lim}, {Pierfederici}, {Pontzen},
  {Ptak}, {Refsdal}, {Servillat}, \& {Streicher}}]{AstropyCollaboration2013}
{Astropy Collaboration}, {Robitaille}, T.~P., {Tollerud}, E.~J., {et~al.} 2013,
  \aap, 558, A33, \dodoi{10.1051/0004-6361/201322068}

\bibitem[{{Astropy Collaboration} {et~al.}(2018){Astropy Collaboration},
  {Price-Whelan}, {Sip{\H{o}}cz}, {G{\"u}nther}, {Lim}, {Crawford}, {Conseil},
  {Shupe}, {Craig}, {Dencheva}, {Ginsburg}, {VanderPlas}, {Bradley},
  {P{\'e}rez-Su{\'a}rez}, {de Val-Borro}, {Aldcroft}, {Cruz}, {Robitaille},
  {Tollerud}, {Ardelean}, {Babej}, {Bach}, {Bachetti}, {Bakanov}, {Bamford},
  {Barentsen}, {Barmby}, {Baumbach}, {Berry}, {Biscani}, {Boquien}, {Bostroem},
  {Bouma}, {Brammer}, {Bray}, {Breytenbach}, {Buddelmeijer}, {Burke},
  {Calderone}, {Cano Rodr{\'\i}guez}, {Cara}, {Cardoso}, {Cheedella}, {Copin},
  {Corrales}, {Crichton}, {D'Avella}, {Deil}, {Depagne}, {Dietrich}, {Donath},
  {Droettboom}, {Earl}, {Erben}, {Fabbro}, {Ferreira}, {Finethy}, {Fox},
  {Garrison}, {Gibbons}, {Goldstein}, {Gommers}, {Greco}, {Greenfield},
  {Groener}, {Grollier}, {Hagen}, {Hirst}, {Homeier}, {Horton}, {Hosseinzadeh},
  {Hu}, {Hunkeler}, {Ivezi{\'c}}, {Jain}, {Jenness}, {Kanarek}, {Kendrew},
  {Kern}, {Kerzendorf}, {Khvalko}, {King}, {Kirkby}, {Kulkarni}, {Kumar},
  {Lee}, {Lenz}, {Littlefair}, {Ma}, {Macleod}, {Mastropietro}, {McCully},
  {Montagnac}, {Morris}, {Mueller}, {Mumford}, {Muna}, {Murphy}, {Nelson},
  {Nguyen}, {Ninan}, {N{\"o}the}, {Ogaz}, {Oh}, {Parejko}, {Parley}, {Pascual},
  {Patil}, {Patil}, {Plunkett}, {Prochaska}, {Rastogi}, {Reddy Janga},
  {Sabater}, {Sakurikar}, {Seifert}, {Sherbert}, {Sherwood-Taylor}, {Shih},
  {Sick}, {Silbiger}, {Singanamalla}, {Singer}, {Sladen}, {Sooley},
  {Sornarajah}, {Streicher}, {Teuben}, {Thomas}, {Tremblay}, {Turner},
  {Terr{\'o}n}, {van Kerkwijk}, {de la Vega}, {Watkins}, {Weaver}, {Whitmore},
  {Woillez}, {Zabalza}, \& {Astropy Contributors}}]{AstropyCollaboration2018}
{Astropy Collaboration}, {Price-Whelan}, A.~M., {Sip{\H{o}}cz}, B.~M., {et~al.}
  2018, \aj, 156, 123, \dodoi{10.3847/1538-3881/aabc4f}

\bibitem[{{Astropy Collaboration} {et~al.}(2022){Astropy Collaboration},
  {Price-Whelan}, {Lim}, {Earl}, {Starkman}, {Bradley}, {Shupe}, {Patil},
  {Corrales}, {Brasseur}, {N{\"o}the}, {Donath}, {Tollerud}, {Morris},
  {Ginsburg}, {Vaher}, {Weaver}, {Tocknell}, {Jamieson}, {van Kerkwijk},
  {Robitaille}, {Merry}, {Bachetti}, {G{\"u}nther}, {Aldcroft},
  {Alvarado-Montes}, {Archibald}, {B{\'o}di}, {Bapat}, {Barentsen},
  {Baz{\'a}n}, {Biswas}, {Boquien}, {Burke}, {Cara}, {Cara}, {Conroy},
  {Conseil}, {Craig}, {Cross}, {Cruz}, {D'Eugenio}, {Dencheva}, {Devillepoix},
  {Dietrich}, {Eigenbrot}, {Erben}, {Ferreira}, {Foreman-Mackey}, {Fox},
  {Freij}, {Garg}, {Geda}, {Glattly}, {Gondhalekar}, {Gordon}, {Grant},
  {Greenfield}, {Groener}, {Guest}, {Gurovich}, {Handberg}, {Hart},
  {Hatfield-Dodds}, {Homeier}, {Hosseinzadeh}, {Jenness}, {Jones}, {Joseph},
  {Kalmbach}, {Karamehmetoglu}, {Ka{\l}uszy{\'n}ski}, {Kelley}, {Kern},
  {Kerzendorf}, {Koch}, {Kulumani}, {Lee}, {Ly}, {Ma}, {MacBride}, {Maljaars},
  {Muna}, {Murphy}, {Norman}, {O'Steen}, {Oman}, {Pacifici}, {Pascual},
  {Pascual-Granado}, {Patil}, {Perren}, {Pickering}, {Rastogi}, {Roulston},
  {Ryan}, {Rykoff}, {Sabater}, {Sakurikar}, {Salgado}, {Sanghi}, {Saunders},
  {Savchenko}, {Schwardt}, {Seifert-Eckert}, {Shih}, {Jain}, {Shukla}, {Sick},
  {Simpson}, {Singanamalla}, {Singer}, {Singhal}, {Sinha}, {Sip{\H{o}}cz},
  {Spitler}, {Stansby}, {Streicher}, {{\v{S}}umak}, {Swinbank}, {Taranu},
  {Tewary}, {Tremblay}, {de Val-Borro}, {Van Kooten}, {Vasovi{\'c}}, {Verma},
  {de Miranda Cardoso}, {Williams}, {Wilson}, {Winkel}, {Wood-Vasey}, {Xue},
  {Yoachim}, {Zhang}, {Zonca}, \& {Astropy Project
  Contributors}}]{AstropyCollaboration2022}
{Astropy Collaboration}, {Price-Whelan}, A.~M., {Lim}, P.~L., {et~al.} 2022,
  \apj, 935, 167, \dodoi{10.3847/1538-4357/ac7c74}

\bibitem[{{Badenas-Agusti} {et~al.}(2024){Badenas-Agusti}, {Via{\~n}a},
  {Vanderburg}, {Blouin}, {Dufour}, {Xu}, \& {Sha}}]{Badenas-Agusti2024}
{Badenas-Agusti}, M., {Via{\~n}a}, J., {Vanderburg}, A., {et~al.} 2024, \mnras,
  529, 1688, \dodoi{10.1093/mnras/stae421}

\bibitem[{{Batalha} {et~al.}(2017){Batalha}, {Mandell}, {Pontoppidan},
  {Stevenson}, {Lewis}, {Kalirai}, {Earl}, {Greene}, {Albert}, \&
  {Nielsen}}]{PandExo}
{Batalha}, N.~E., {Mandell}, A., {Pontoppidan}, K., {et~al.} 2017, \pasp, 129,
  064501, \dodoi{10.1088/1538-3873/aa65b0}

\bibitem[{{Bean} {et~al.}(2018){Bean}, {Stevenson}, {Batalha},
  {Berta-Thompson}, {Kreidberg}, {Crouzet}, {Benneke}, {Line}, {Sing},
  {Wakeford}, {Knutson}, {Kempton}, {D{\'e}sert}, {Crossfield}, {Batalha}, {de
  Wit}, {Parmentier}, {Harrington}, {Moses}, {Lopez-Morales}, {Alam}, {Blecic},
  {Bruno}, {Carter}, {Chapman}, {Decin}, {Dragomir}, {Evans}, {Fortney},
  {Fraine}, {Gao}, {Garc{\'\i}a Mu{\~n}oz}, {Gibson}, {Goyal}, {Heng}, {Hu},
  {Kendrew}, {Kilpatrick}, {Krick}, {Lagage}, {Lendl}, {Louden}, {Madhusudhan},
  {Mandell}, {Mansfield}, {May}, {Morello}, {Morley}, {Nikolov}, {Redfield},
  {Roberts}, {Schlawin}, {Spake}, {Todorov}, {Tsiaras}, {Venot}, {Waalkes},
  {Wheatley}, {Zellem}, {Angerhausen}, {Barrado}, {Carone}, {Casewell},
  {Cubillos}, {Damiano}, {de Val-Borro}, {Drummond}, {Edwards}, {Endl},
  {Espinoza}, {France}, {Gizis}, {Greene}, {Henning}, {Hong}, {Ingalls}, {Iro},
  {Irwin}, {Kataria}, {Lahuis}, {Leconte}, {Lillo-Box}, {Lines}, {Lothringer},
  {Mancini}, {Marchis}, {Mayne}, {Palle}, {Rauscher}, {Roudier}, {Shkolnik},
  {Southworth}, {Swain}, {Taylor}, {Teske}, {Tinetti}, {Tremblin}, {Tucker},
  {van Boekel}, {Waldmann}, {Weaver}, \& {Zingales}}]{Bean2018}
{Bean}, J.~L., {Stevenson}, K.~B., {Batalha}, N.~M., {et~al.} 2018, \pasp, 130,
  114402, \dodoi{10.1088/1538-3873/aadbf3}

\bibitem[{{Benneke} \& {Seager}(2013)}]{Benneke2013}
{Benneke}, B., \& {Seager}, S. 2013, \apj, 778, 153,
  \dodoi{10.1088/0004-637X/778/2/153}

\bibitem[{{Berardo} {et~al.}(2024){Berardo}, {de Wit}, \&
  {Rackham}}]{Berardo2024}
{Berardo}, D., {de Wit}, J., \& {Rackham}, B.~V. 2024, \apjl, 961, L18,
  \dodoi{10.3847/2041-8213/ad1b5b}

\bibitem[{{Brown}(2001)}]{Brown2001}
{Brown}, T.~M. 2001, \apj, 553, 1006, \dodoi{10.1086/320950}

\bibitem[{{Buchner}(2014)}]{Buchner2014}
{Buchner}, J. 2014, arXiv e-prints, arXiv:1407.5459,
  \dodoi{10.48550/arXiv.1407.5459}

\bibitem[{{Buchner}(2017)}]{Buchner2017}
---. 2017, arXiv e-prints, arXiv:1707.04476, \dodoi{10.48550/arXiv.1707.04476}

\bibitem[{{Buchner}(2021)}]{Buchner2021}
---. 2021, The Journal of Open Source Software, 6, 3001,
  \dodoi{10.21105/joss.03001}

\bibitem[{{Czekala} {et~al.}(2015){Czekala}, {Andrews}, {Mandel}, {Hogg}, \&
  {Green}}]{Czekala2015}
{Czekala}, I., {Andrews}, S.~M., {Mandel}, K.~S., {Hogg}, D.~W., \& {Green},
  G.~M. 2015, \apj, 812, 128, \dodoi{10.1088/0004-637X/812/2/128}

\bibitem[{{de Wit} {et~al.}(2012){de Wit}, {Gillon}, {Demory}, \&
  {Seager}}]{deWit2012}
{de Wit}, J., {Gillon}, M., {Demory}, B.~O., \& {Seager}, S. 2012, \aap, 548,
  A128, \dodoi{10.1051/0004-6361/201219060}

\bibitem[{{Espinoza} {et~al.}(2019){Espinoza}, {Rackham}, {Jord{\'a}n}, {Apai},
  {L{\'o}pez-Morales}, {Osip}, {Grimm}, {Hoeijmakers}, {Wilson}, {Bixel},
  {McGruder}, {Rodler}, {Weaver}, {Lewis}, {Fortney}, \&
  {Fraine}}]{Espinoza2019}
{Espinoza}, N., {Rackham}, B.~V., {Jord{\'a}n}, A., {et~al.} 2019, \mnras, 482,
  2065, \dodoi{10.1093/mnras/sty2691}

\bibitem[{{Feinstein} {et~al.}(2023){Feinstein}, {Radica}, {Welbanks},
  {Murray}, {Ohno}, {Coulombe}, {Espinoza}, {Bean}, {Teske}, {Benneke}, {Line},
  {Rustamkulov}, {Saba}, {Tsiaras}, {Barstow}, {Fortney}, {Gao}, {Knutson},
  {MacDonald}, {Mikal-Evans}, {Rackham}, {Taylor}, {Parmentier}, {Batalha},
  {Berta-Thompson}, {Carter}, {Changeat}, {dos Santos}, {Gibson}, {Goyal},
  {Kreidberg}, {L{\'o}pez-Morales}, {Lothringer}, {Miguel}, {Molaverdikhani},
  {Moran}, {Morello}, {Mukherjee}, {Sing}, {Stevenson}, {Wakeford}, {Ahrer},
  {Alam}, {Alderson}, {Allen}, {Batalha}, {Bell}, {Blecic}, {Brande},
  {Caceres}, {Casewell}, {Chubb}, {Crossfield}, {Crouzet}, {Cubillos}, {Decin},
  {D{\'e}sert}, {Harrington}, {Heng}, {Henning}, {Iro}, {Kempton}, {Kendrew},
  {Kirk}, {Krick}, {Lagage}, {Lendl}, {Mancini}, {Mansfield}, {May}, {Mayne},
  {Nikolov}, {Palle}, {Petit dit de la Roche}, {Piaulet}, {Powell}, {Redfield},
  {Rogers}, {Roman}, {Roy}, {Nixon}, {Schlawin}, {Tan}, {Tremblin}, {Turner},
  {Venot}, {Waalkes}, {Wheatley}, \& {Zhang}}]{Feinstein2023}
{Feinstein}, A.~D., {Radica}, M., {Welbanks}, L., {et~al.} 2023, \nat, 614,
  670, \dodoi{10.1038/s41586-022-05674-1}

\bibitem[{{Foreman-Mackey} {et~al.}(2013){Foreman-Mackey}, {Hogg}, {Lang}, \&
  {Goodman}}]{emcee}
{Foreman-Mackey}, D., {Hogg}, D.~W., {Lang}, D., \& {Goodman}, J. 2013, \pasp,
  125, 306, \dodoi{10.1086/670067}

\bibitem[{{Foreman-Mackey} {et~al.}(2019){Foreman-Mackey}, {Farr}, {Sinha},
  {Archibald}, {Hogg}, {Sanders}, {Zuntz}, {Williams}, {Nelson}, {de
  Val-Borro}, {Erhardt}, {Pashchenko}, \& {Pla}}]{emcee_v3}
{Foreman-Mackey}, D., {Farr}, W., {Sinha}, M., {et~al.} 2019, The Journal of
  Open Source Software, 4, 1864, \dodoi{10.21105/joss.01864}

\bibitem[{{Fu} {et~al.}(2022){Fu}, {Espinoza}, {Sing}, {Lothringer}, {Dos
  Santos}, {Rustamkulov}, {Deming}, {Kempton}, {Komacek}, {Knutson}, {Albert},
  {Pontoppidan}, {Volk}, \& {Filippazzo}}]{Fu2022}
{Fu}, G., {Espinoza}, N., {Sing}, D.~K., {et~al.} 2022, \apjl, 940, L35,
  \dodoi{10.3847/2041-8213/ac9977}

\bibitem[{{Garcia} {et~al.}(2022){Garcia}, {Moran}, {Rackham}, {Wakeford},
  {Gillon}, {de Wit}, \& {Lewis}}]{Garcia2022}
{Garcia}, L.~J., {Moran}, S.~E., {Rackham}, B.~V., {et~al.} 2022, \aap, 665,
  A19, \dodoi{10.1051/0004-6361/202142603}

\bibitem[{{Gillon} {et~al.}(2016){Gillon}, {Jehin}, {Lederer}, {Delrez}, {de
  Wit}, {Burdanov}, {Van Grootel}, {Burgasser}, {Triaud}, {Opitom}, {Demory},
  {Sahu}, {Bardalez Gagliuffi}, {Magain}, \& {Queloz}}]{Gillon2016}
{Gillon}, M., {Jehin}, E., {Lederer}, S.~M., {et~al.} 2016, \nat, 533, 221,
  \dodoi{10.1038/nature17448}

\bibitem[{{Gillon} {et~al.}(2017){Gillon}, {Triaud}, {Demory}, {Jehin}, {Agol},
  {Deck}, {Lederer}, {de Wit}, {Burdanov}, {Ingalls}, {Bolmont}, {Leconte},
  {Raymond}, {Selsis}, {Turbet}, {Barkaoui}, {Burgasser}, {Burleigh}, {Carey},
  {Chaushev}, {Copperwheat}, {Delrez}, {Fernand es}, {Holdsworth}, {Kotze},
  {Van Grootel}, {Almleaky}, {Benkhaldoun}, {Magain}, \& {Queloz}}]{Gillon2017}
{Gillon}, M., {Triaud}, A. H.~M.~J., {Demory}, B.-O., {et~al.} 2017, \nat, 542,
  456, \dodoi{10.1038/nature21360}

\bibitem[{{Gordon} {et~al.}(2017){Gordon}, {Rothman}, {Hill}, {Kochanov},
  {Tan}, {Bernath}, {Birk}, {Boudon}, {Campargue}, {Chance}, {Drouin}, {Flaud},
  {Gamache}, {Hodges}, {Jacquemart}, {Perevalov}, {Perrin}, {Shine}, {Smith},
  {Tennyson}, {Toon}, {Tran}, {Tyuterev}, {Barbe}, {Cs{\'a}sz{\'a}r}, {Devi},
  {Furtenbacher}, {Harrison}, {Hartmann}, {Jolly}, {Johnson}, {Karman},
  {Kleiner}, {Kyuberis}, {Loos}, {Lyulin}, {Massie}, {Mikhailenko},
  {Moazzen-Ahmadi}, {M{\"u}ller}, {Naumenko}, {Nikitin}, {Polyansky}, {Rey},
  {Rotger}, {Sharpe}, {Sung}, {Starikova}, {Tashkun}, {Auwera}, {Wagner},
  {Wilzewski}, {Wcis{\l}o}, {Yu}, \& {Zak}}]{Gordon2017}
{Gordon}, I.~E., {Rothman}, L.~S., {Hill}, C., {et~al.} 2017, \jqsrt, 203, 3,
  \dodoi{10.1016/j.jqsrt.2017.06.038}

\bibitem[{{Gordon} {et~al.}(2022){Gordon}, {Rothman}, {Hargreaves}, {Hashemi},
  {Karlovets}, {Skinner}, {Conway}, {Hill}, {Kochanov}, {Tan}, {Wcis{\l}o},
  {Finenko}, {Nelson}, {Bernath}, {Birk}, {Boudon}, {Campargue}, {Chance},
  {Coustenis}, {Drouin}, {Flaud}, {Gamache}, {Hodges}, {Jacquemart}, {Mlawer},
  {Nikitin}, {Perevalov}, {Rotger}, {Tennyson}, {Toon}, {Tran}, {Tyuterev},
  {Adkins}, {Baker}, {Barbe}, {Can{\`e}}, {Cs{\'a}sz{\'a}r}, {Dudaryonok},
  {Egorov}, {Fleisher}, {Fleurbaey}, {Foltynowicz}, {Furtenbacher}, {Harrison},
  {Hartmann}, {Horneman}, {Huang}, {Karman}, {Karns}, {Kassi}, {Kleiner},
  {Kofman}, {Kwabia-Tchana}, {Lavrentieva}, {Lee}, {Long}, {Lukashevskaya},
  {Lyulin}, {Makhnev}, {Matt}, {Massie}, {Melosso}, {Mikhailenko}, {Mondelain},
  {M{\"u}ller}, {Naumenko}, {Perrin}, {Polyansky}, {Raddaoui}, {Raston},
  {Reed}, {Rey}, {Richard}, {T{\'o}bi{\'a}s}, {Sadiek}, {Schwenke},
  {Starikova}, {Sung}, {Tamassia}, {Tashkun}, {Vander Auwera}, {Vasilenko},
  {Vigasin}, {Villanueva}, {Vispoel}, {Wagner}, {Yachmenev}, \&
  {Yurchenko}}]{Gordon2022}
{Gordon}, I.~E., {Rothman}, L.~S., {Hargreaves}, R.~J., {et~al.} 2022, \jqsrt,
  277, 107949, \dodoi{10.1016/j.jqsrt.2021.107949}

\bibitem[{Harris {et~al.}(2020)Harris, Millman, van~der Walt, Gommers,
  Virtanen, Cournapeau, Wieser, Taylor, Berg, Smith, Kern, Picus, Hoyer, van
  Kerkwijk, Brett, Haldane, del R{\'{i}}o, Wiebe, Peterson,
  G{\'{e}}rard-Marchant, Sheppard, Reddy, Weckesser, Abbasi, Gohlke, \&
  Oliphant}]{NumPy}
Harris, C.~R., Millman, K.~J., van~der Walt, S.~J., {et~al.} 2020, Nature, 585,
  357, \dodoi{10.1038/s41586-020-2649-2}

\bibitem[{{Hunter}(2007)}]{Hunter2007}
{Hunter}, J.~D. 2007, Computing in Science and Engineering, 9, 90,
  \dodoi{10.1109/MCSE.2007.55}

\bibitem[{{Husser} {et~al.}(2013){Husser}, {Wende-von Berg}, {Dreizler},
  {Homeier}, {Reiners}, {Barman}, \& {Hauschildt}}]{Husser2013}
{Husser}, T.~O., {Wende-von Berg}, S., {Dreizler}, S., {et~al.} 2013, \aap,
  553, A6, \dodoi{10.1051/0004-6361/201219058}

\bibitem[{{Iyer} \& {Line}(2020)}]{Iyer2020}
{Iyer}, A.~R., \& {Line}, M.~R. 2020, \apj, 889, 78,
  \dodoi{10.3847/1538-4357/ab612e}

\bibitem[{{Iyer} {et~al.}(2023){Iyer}, {Line}, {Muirhead}, {Fortney}, \&
  {Gharib-Nezhad}}]{Iyer2023}
{Iyer}, A.~R., {Line}, M.~R., {Muirhead}, P.~S., {Fortney}, J.~J., \&
  {Gharib-Nezhad}, E. 2023, \apj, 944, 41, \dodoi{10.3847/1538-4357/acabc2}

\bibitem[{{JWST Transiting Exoplanet Community Early Release Science Team}
  {et~al.}(2023){JWST Transiting Exoplanet Community Early Release Science
  Team}, {Ahrer}, {Alderson}, {Batalha}, {Batalha}, {Bean}, {Beatty}, {Bell},
  {Benneke}, {Berta-Thompson}, {Carter}, {Crossfield}, {Espinoza}, {Feinstein},
  {Fortney}, {Gibson}, {Goyal}, {Kempton}, {Kirk}, {Kreidberg},
  {L{\'o}pez-Morales}, {Line}, {Lothringer}, {Moran}, {Mukherjee}, {Ohno},
  {Parmentier}, {Piaulet}, {Rustamkulov}, {Schlawin}, {Sing}, {Stevenson},
  {Wakeford}, {Allen}, {Birkmann}, {Brande}, {Crouzet}, {Cubillos}, {Damiano},
  {D{\'e}sert}, {Gao}, {Harrington}, {Hu}, {Kendrew}, {Knutson}, {Lagage},
  {Leconte}, {Lendl}, {MacDonald}, {May}, {Miguel}, {Molaverdikhani}, {Moses},
  {Murray}, {Nehring}, {Nikolov}, {Petit dit de la Roche}, {Radica}, {Roy},
  {Stassun}, {Taylor}, {Waalkes}, {Wachiraphan}, {Welbanks}, {Wheatley},
  {Aggarwal}, {Alam}, {Banerjee}, {Barstow}, {Blecic}, {Casewell}, {Changeat},
  {Chubb}, {Col{\'o}n}, {Coulombe}, {Daylan}, {de Val-Borro}, {Decin}, {Dos
  Santos}, {Flagg}, {France}, {Fu}, {Garc{\'\i}a Mu{\~n}oz}, {Gizis},
  {Glidden}, {Grant}, {Heng}, {Henning}, {Hong}, {Inglis}, {Iro}, {Kataria},
  {Komacek}, {Krick}, {Lee}, {Lewis}, {Lillo-Box}, {Lustig-Yaeger}, {Mancini},
  {Mandell}, {Mansfield}, {Marley}, {Mikal-Evans}, {Morello}, {Nixon}, {Ortiz
  Ceballos}, {Piette}, {Powell}, {Rackham}, {Ramos-Rosado}, {Rauscher},
  {Redfield}, {Rogers}, {Roman}, {Roudier}, {Scarsdale}, {Shkolnik},
  {Southworth}, {Spake}, {Steinrueck}, {Tan}, {Teske}, {Tremblin}, {Tsai},
  {Tucker}, {Turner}, {Valenti}, {Venot}, {Waldmann}, {Wallack}, {Zhang}, \&
  {Zieba}}]{JTECERST2023}
{JWST Transiting Exoplanet Community Early Release Science Team}, {Ahrer},
  E.-M., {Alderson}, L., {et~al.} 2023, \nat, 614, 649,
  \dodoi{10.1038/s41586-022-05269-w}

\bibitem[{{Kostogryz} {et~al.}(2023){Kostogryz}, {Shapiro}, {Witzke}, {Grant},
  {Wakeford}, {Stevenson}, {Solanki}, \& {Gizon}}]{Kostogryz2023}
{Kostogryz}, N., {Shapiro}, A.~I., {Witzke}, V., {et~al.} 2023, Research Notes
  of the American Astronomical Society, 7, 39, \dodoi{10.3847/2515-5172/acc180}

\bibitem[{{Lim} {et~al.}(2023){Lim}, {Benneke}, {Doyon}, {MacDonald},
  {Piaulet}, {Artigau}, {Coulombe}, {Radica}, {L'Heureux}, {Albert}, {Rackham},
  {de Wit}, {Salhi}, {Roy}, {Flagg}, {Fournier-Tondreau}, {Taylor}, {Cook},
  {Lafreni{\`e}re}, {Cowan}, {Kaltenegger}, {Rowe}, {Espinoza}, {Dang}, \&
  {Darveau-Bernier}}]{Lim2023}
{Lim}, O., {Benneke}, B., {Doyon}, R., {et~al.} 2023, \apjl, 955, L22,
  \dodoi{10.3847/2041-8213/acf7c4}

\bibitem[{{McCullough} {et~al.}(2014){McCullough}, {Crouzet}, {Deming}, \&
  {Madhusudhan}}]{McCullough2014}
{McCullough}, P.~R., {Crouzet}, N., {Deming}, D., \& {Madhusudhan}, N. 2014,
  \apj, 791, 55, \dodoi{10.1088/0004-637X/791/1/55}

\bibitem[{{M}c{K}inney(2010)}]{pandas}
{M}c{K}inney, W. 2010, in {P}roceedings of the 9th {P}ython in {S}cience
  {C}onference, ed. {S}t\'efan van~der {W}alt \& {J}arrod {M}illman, 56 -- 61,
  \dodoi{10.25080/Majora-92bf1922-00a}

\bibitem[{{Niraula} {et~al.}(2022){Niraula}, {de Wit}, {Gordon}, {Hargreaves},
  {Sousa-Silva}, \& {Kochanov}}]{Niraula2022}
{Niraula}, P., {de Wit}, J., {Gordon}, I.~E., {et~al.} 2022, Nature Astronomy,
  \dodoi{10.1038/s41550-022-01773-1}

\bibitem[{{Pecaut} \& {Mamajek}(2013)}]{Pecaut2013}
{Pecaut}, M.~J., \& {Mamajek}, E.~E. 2013, \apjs, 208, 9,
  \dodoi{10.1088/0067-0049/208/1/9}

\bibitem[{{Pinhas} {et~al.}(2018){Pinhas}, {Rackham}, {Madhusudhan}, \&
  {Apai}}]{Pinhas2018}
{Pinhas}, A., {Rackham}, B.~V., {Madhusudhan}, N., \& {Apai}, D. 2018, \mnras,
  480, 5314, \dodoi{10.1093/mnras/sty2209}

\bibitem[{{Rackham} {et~al.}(2017){Rackham}, {Espinoza}, {Apai},
  {L{\'o}pez-Morales}, {Jord{\'a}n}, {Osip}, {Lewis}, {Rodler}, {Fraine},
  {Morley}, \& {Fortney}}]{Rackham2017}
{Rackham}, B., {Espinoza}, N., {Apai}, D., {et~al.} 2017, \apj, 834, 151,
  \dodoi{10.3847/1538-4357/aa4f6c}

\bibitem[{{Rackham}(2023)}]{speclib-0.0-beta.0}
{Rackham}, B.~V. 2023, {speclib}, 0.0-beta.0, Zenodo,  Zenodo,
  \dodoi{10.5281/zenodo.7868050}

\bibitem[{{Rackham} {et~al.}(2018){Rackham}, {Apai}, \&
  {Giampapa}}]{Rackham2018}
{Rackham}, B.~V., {Apai}, D., \& {Giampapa}, M.~S. 2018, \apj, 853, 122,
  \dodoi{10.3847/1538-4357/aaa08c}

\bibitem[{{Rackham} {et~al.}(2019){Rackham}, {Apai}, \&
  {Giampapa}}]{Rackham2019}
---. 2019, \aj, 157, 96, \dodoi{10.3847/1538-3881/aaf892}

\bibitem[{{Rackham} {et~al.}(2023){Rackham}, {Espinoza}, {Berdyugina},
  {Korhonen}, {MacDonald}, {Montet}, {Morris}, {Oshagh}, {Shapiro}, {Unruh},
  {Quintana}, {Zellem}, {Apai}, {Barclay}, {Barstow}, {Bruno}, {Carone},
  {Casewell}, {Cegla}, {Criscuoli}, {Fischer}, {Fournier}, {Giampapa}, {Giles},
  {Iyer}, {Kopp}, {Kostogryz}, {Krivova}, {Mallonn}, {McGruder},
  {Molaverdikhani}, {Newton}, {Panja}, {Peacock}, {Reardon}, {Roettenbacher},
  {Scandariato}, {Solanki}, {Stassun}, {Steiner}, {Stevenson}, {Tregloan-Reed},
  {Valio}, {Wedemeyer}, {Welbanks}, {Yu}, {Alam}, {Davenport}, {Deming},
  {Dong}, {Ducrot}, {Fisher}, {Gilbert}, {Kostov}, {L{\'o}pez-Morales}, {Line},
  {Mo{\v{c}}nik}, {Mullally}, {Paudel}, {Ribas}, \& {Valenti}}]{Rackham2023}
{Rackham}, B.~V., {Espinoza}, N., {Berdyugina}, S.~V., {et~al.} 2023, RAS
  Techniques and Instruments, 2, 148, \dodoi{10.1093/rasti/rzad009}

\bibitem[{{Rustamkulov} {et~al.}(2022){Rustamkulov}, {Sing}, {Liu}, \&
  {Wang}}]{Rustamkulov2022}
{Rustamkulov}, Z., {Sing}, D.~K., {Liu}, R., \& {Wang}, A. 2022, \apjl, 928,
  L7, \dodoi{10.3847/2041-8213/ac5b6f}

\bibitem[{{Rustamkulov} {et~al.}(2023){Rustamkulov}, {Sing}, {Mukherjee},
  {May}, {Kirk}, {Schlawin}, {Line}, {Piaulet}, {Carter}, {Batalha}, {Goyal},
  {L{\'o}pez-Morales}, {Lothringer}, {MacDonald}, {Moran}, {Stevenson},
  {Wakeford}, {Espinoza}, {Bean}, {Batalha}, {Benneke}, {Berta-Thompson},
  {Crossfield}, {Gao}, {Kreidberg}, {Powell}, {Cubillos}, {Gibson}, {Leconte},
  {Molaverdikhani}, {Nikolov}, {Parmentier}, {Roy}, {Taylor}, {Turner},
  {Wheatley}, {Aggarwal}, {Ahrer}, {Alam}, {Alderson}, {Allen}, {Banerjee},
  {Barat}, {Barrado}, {Barstow}, {Bell}, {Blecic}, {Brande}, {Casewell},
  {Changeat}, {Chubb}, {Crouzet}, {Daylan}, {Decin}, {D{\'e}sert},
  {Mikal-Evans}, {Feinstein}, {Flagg}, {Fortney}, {Harrington}, {Heng}, {Hong},
  {Hu}, {Iro}, {Kataria}, {Kempton}, {Krick}, {Lendl}, {Lillo-Box}, {Louca},
  {Lustig-Yaeger}, {Mancini}, {Mansfield}, {Mayne}, {Miguel}, {Morello},
  {Ohno}, {Palle}, {Petit dit de la Roche}, {Rackham}, {Radica},
  {Ramos-Rosado}, {Redfield}, {Rogers}, {Shkolnik}, {Southworth}, {Teske},
  {Tremblin}, {Tucker}, {Venot}, {Waalkes}, {Welbanks}, {Zhang}, \&
  {Zieba}}]{Rustamkulov2023_ERS}
{Rustamkulov}, Z., {Sing}, D.~K., {Mukherjee}, S., {et~al.} 2023, \nat, 614,
  659, \dodoi{10.1038/s41586-022-05677-y}

\bibitem[{{Seager} \& {Sasselov}(2000)}]{Seager2000}
{Seager}, S., \& {Sasselov}, D.~D. 2000, \apj, 537, 916, \dodoi{10.1086/309088}

\bibitem[{{Sing} {et~al.}(2011){Sing}, {Pont}, {Aigrain}, {Charbonneau},
  {D{\'e}sert}, {Gibson}, {Gilliland}, {Hayek}, {Henry}, {Knutson}, {Lecavelier
  Des Etangs}, {Mazeh}, \& {Shporer}}]{Sing2011}
{Sing}, D.~K., {Pont}, F., {Aigrain}, S., {et~al.} 2011, \mnras, 416, 1443,
  \dodoi{10.1111/j.1365-2966.2011.19142.x}

\bibitem[{{Tennyson} {et~al.}(2016){Tennyson}, {Yurchenko}, {Al-Refaie},
  {Barton}, {Chubb}, {Coles}, {Diamantopoulou}, {Gorman}, {Hill}, {Lam},
  {Lodi}, {McKemmish}, {Na}, {Owens}, {Polyansky}, {Rivlin}, {Sousa-Silva},
  {Underwood}, {Yachmenev}, \& {Zak}}]{Tennyson2016}
{Tennyson}, J., {Yurchenko}, S.~N., {Al-Refaie}, A.~F., {et~al.} 2016, Journal
  of Molecular Spectroscopy, 327, 73, \dodoi{10.1016/j.jms.2016.05.002}

\bibitem[{{Tennyson} {et~al.}(2020){Tennyson}, {Yurchenko}, {Al-Refaie},
  {Clark}, {Chubb}, {Conway}, {Dewan}, {Gorman}, {Hill}, {Lynas-Gray},
  {Mellor}, {McKemmish}, {Owens}, {Polyansky}, {Semenov}, {Somogyi}, {Tinetti},
  {Upadhyay}, {Waldmann}, {Wang}, {Wright}, \& {Yurchenko}}]{Tennyson2020}
---. 2020, \jqsrt, 255, 107228, \dodoi{10.1016/j.jqsrt.2020.107228}

\bibitem[{{TRAPPIST-1 JWST Community Initiative} {et~al.}(2023){TRAPPIST-1 JWST
  Community Initiative}, {de Wit}, {Doyon}, {Rackham}, {Lim}, {Ducrot},
  {Kreidberg}, {Benneke}, {Ribas}, {Berardo}, {Niraula}, {Iyer}, {Shapiro},
  {Kostogryz}, {Witzke}, {Gillon}, {Agol}, {Meadows}, {Burgasser}, {Owen},
  {Fortney}, {Selsis}, {Bello-Arufe}, {Bolmont}, {Cowan}, {Dong}, {Drake},
  {Garcia}, {Greene}, {Haworth}, {Hu}, {Kane}, {Kervella}, {Koll},
  {Krissansen-Totton}, {Lagage}, {Lichtenberg}, {Lustig-Yaeger}, {Lingam},
  {Turbet}, {Seager}, {Barkaoui}, {Bell}, {Burdanov}, {Cadieux}, {Charnay},
  {Cloutier}, {Cook}, {Correia}, {Dang}, {Daylan}, {Delrez}, {Edwards},
  {Fauchez}, {Flagg}, {Fraschetti}, {Haqq-Misra}, {Huang}, {Iro},
  {Jayawardhana}, {Jehin}, {Jin}, {Kite}, {Kitzmann}, {Kral}, {Lafreni{\`e}re},
  {Libert}, {Liu}, {Mohanty}, {Morris}, {Murray}, {Piaulet}, {Pozuelos},
  {Radica}, {Ranjan}, {Rathcke}, {Roy}, {Schwieterman}, {Turner}, {Triaud}, \&
  {Way}}]{TRAPPIST-1JWSTCommunityInitiative2023}
{TRAPPIST-1 JWST Community Initiative}, {de Wit}, J., {Doyon}, R., {et~al.}
  2023, arXiv e-prints, arXiv:2310.15895, \dodoi{10.48550/arXiv.2310.15895}

\bibitem[{{Trotta}(2008)}]{Trotta2008}
{Trotta}, R. 2008, Contemporary Physics, 49, 71,
  \dodoi{10.1080/00107510802066753}

\bibitem[{Virtanen {et~al.}(2020)Virtanen, Gommers, Oliphant, Haberland, Reddy,
  Cournapeau, Burovski, Peterson, Weckesser, Bright, {van der Walt}, Brett,
  Wilson, Millman, Mayorov, Nelson, Jones, Kern, Larson, Carey, Polat, Feng,
  Moore, {VanderPlas}, Laxalde, Perktold, Cimrman, Henriksen, Quintero, Harris,
  Archibald, Ribeiro, Pedregosa, {van Mulbregt}, \& {SciPy 1.0
  Contributors}}]{SciPy}
Virtanen, P., Gommers, R., Oliphant, T.~E., {et~al.} 2020, Nature Methods, 17,
  261, \dodoi{10.1038/s41592-019-0686-2}

\bibitem[{{Wakeford} {et~al.}(2019){Wakeford}, {Lewis}, {Fowler}, {Bruno},
  {Wilson}, {Moran}, {Valenti}, {Batalha}, {Filippazzo}, {Bourrier},
  {H{\"o}rst}, {Lederer}, \& {de Wit}}]{Wakeford2019}
{Wakeford}, H.~R., {Lewis}, N.~K., {Fowler}, J., {et~al.} 2019, \aj, 157, 11,
  \dodoi{10.3847/1538-3881/aaf04d}

\bibitem[{{Wenger} {et~al.}(2000){Wenger}, {Ochsenbein}, {Egret}, {Dubois},
  {Bonnarel}, {Borde}, {Genova}, {Jasniewicz}, {Lalo{\"e}}, {Lesteven}, \&
  {Monier}}]{Simbad}
{Wenger}, M., {Ochsenbein}, F., {Egret}, D., {et~al.} 2000, \aaps, 143, 9,
  \dodoi{10.1051/aas:2000332}

\bibitem[{{Witzke} {et~al.}(2022){Witzke}, {Shapiro}, {Kostogryz}, {Cameron},
  {Rackham}, {Seager}, {Solanki}, \& {Unruh}}]{Witzke2022}
{Witzke}, V., {Shapiro}, A.~I., {Kostogryz}, N.~M., {et~al.} 2022, \apjl, 941,
  L35, \dodoi{10.3847/2041-8213/aca671}

\bibitem[{{Witzke} {et~al.}(2021){Witzke}, {Shapiro}, {Cernetic}, {Tagirov},
  {Kostogryz}, {Anusha}, {Unruh}, {Solanki}, \& {Kurucz}}]{Witzke2021}
{Witzke}, V., {Shapiro}, A.~I., {Cernetic}, M., {et~al.} 2021, \aap, 653, A65,
  \dodoi{10.1051/0004-6361/202140275}

\bibitem[{{Zhang} {et~al.}(2018){Zhang}, {Zhou}, {Rackham}, \&
  {Apai}}]{ZhangZhanbo2018}
{Zhang}, Z., {Zhou}, Y., {Rackham}, B.~V., \& {Apai}, D. 2018, \aj, 156, 178,
  \dodoi{10.3847/1538-3881/aade4f}

\end{thebibliography}
\bibliographystyle{aasjournal}



\end{document}